# Optimizing Sample Size for Supervised Machine Learning with Bulk Transcriptomic Sequencing: A Learning Curve Approach

Yunhui Qi[1,2], Xinyi Wang[1,3], Li-Xuan Qin[1,*]

[1] Department of Epidemiology and Biostatistics, Memorial Sloan Kettering Cancer Center, New York, NY, United States

[2] Department of Statistics, Iowa State University, Ames, IA, United States

[3] Department of Statistics, The University of California, Davis, CA, United States

* Corresponding Author: qinl@mskcc.org

# Abstract

Accurate sample classification using transcriptomics data is crucial for advancing personalized medicine. Achieving this goal necessitates determining a suitable sample size that ensures adequate statistical power without undue resource allocation. Current sample size calculation methods rely on assumptions and algorithms that may not align with supervised machine learning techniques for sample classification. Addressing this critical methodological gap, we present a novel computational approach that establishes the power-versus-sample-size relationship by employing a data augmentation strategy followed by fitting a learning curve. We comprehensively evaluated its performance for microRNA and RNA sequencing data, considering diverse data characteristics and algorithm configurations, based on a spectrum of evaluation metrics. To foster accessibility and reproducibility, the Python and R code for implementing our approach is available on GitHub. Its deployment will significantly facilitate the adoption of machine learning in transcriptomics studies and accelerate their translation into clinically useful classifiers for personalized treatment.

# Introduction

Accurate sample classification using transcriptomic sequencing data is pivotal for guiding personalized treatment decisions [1-6]. The success of such endeavors depends on the selection of an appropriate sample size, to achieve adequate statistical power while avoiding undue resource allocation or ethical concerns [7-12]. Various sample size calculation methods are available to identify differentially expressed markers [13-19]. These methods establish connections between the required sample size, the desired power, and the projected effect size within a hypothesis testing framework, employing either closed-form formulae derived from statistical tests [13-16] or *in silico* simulations based on parametric distributions [17-19]. When the study goal shifts to developing a multi-marker classifier, fewer sample size calculation methods are available. They were primarily developed for microarray data and, in principle, can be adapted for sequencing data [20-24]. These methods establish relationships between the required sample size and the desired classification accuracy, through either formulae derived from parametric distributions [20-22] or simulations via subsampling [23,24]. However, none of these methods are compatible with modern supervised machine learning techniques, as these techniques eschew parametric distribution assumptions and require a substantial number of samples, making subsampling infeasible [25-28]. Consequently, there is a pressing need to develop sample size calculation methods compatible with machine learning in classification studies using transcriptomic sequencing data.

We developed a new computational approach to fill this methodology gap. Our approach entails two stages: first, synthesizing realistically distributed transcriptomic sequencing data without relying on a predefined formula, and second, determining a suitable sample size based on the synthesized data across a range of sample sizes. Specifically, we (1) build data augmentation tools that harness the power of deep generative models (DGMs), which will be trained on available pilot data and subsequently used to generate data for any desired number of samples [29-31], and (2) ascertain a suitable sample size by fitting the inverse power law function (IPLF) with augmented data across different sample sizes and their respective classification accuracies using a machine learning technique [32,33] (Fig. 1). We name our algorithm for the first stage SyNG-BTS (pronounced 'sing-beats'), representing Synthesis of Next Generation Bulk Transcriptomic Sequencing, and the algorithm for the second stage SyntheSize.

DGMs are designed to simulate data resembling real-world observations, which can be especially useful when acquiring real data is challenging [34-37]. DGMs initially received acclaim for augmenting imaging data [34] and recently achieved successes in single cell sequencing [35-37]. Several families of DGMs are popularly used for data augmentation, including Variational Autoencoders (VAEs), Generative Adversarial Networks (GANs), and Flow-based generative models [38-41]. When employing these models to simulate bulk-tissue transcriptomic sequencing, a challenge arises due to the typically modest sample size of the training data, as its randomness after online augmentation (that is, augmentation by DGMs in real-time) may have a strong impact on the training process. To address this issue, we opt for offline augmentation, which generates and stores augmented copies of the original data before training, ensuring a more consistent set of generated samples and fostering more stable learning dynamics. We will utilize an autoencoder (AE) for offline augmentation when dealing with a relatively modest number of markers, such as in microRNA sequencing (miRNA-seq) [42-45]. AEs are designed for reconstructing the input data rather than generating new samples. In the case of RNA sequencing (RNA-seq), where the number of markers is substantial, we will employ Gaussian white noise addition [46,47]. It introduces noise generated from a Gaussian distribution to the pilot data and iterates this procedure multiple times. The resulting datasets are aggregated to expand the sample size of the original pilot data.

IPLFs are utilized to represent learning curves that depict the relationship between a classifier's accuracy and the training data's sample size [32,33]. The term "learning curve" is often used to portray the learning trajectory, illustrating how a learner's performance improves with experience and practice [33]. In the realm of machine learning, a learning curve refers to a graphical representation or a mathematical function that illustrates how a learning model's performance improves with an increasing amount of training data. Such curves establish a connection between the prediction accuracy for a learning technique and the sample size of the training dataset. More specifically, these curves typically adhere to the IPLF, displaying three distinct and sequential phases characterized by (1) rapid performance enhancement, (2) gradual progress, and (3) eventual plateauing. The IPLF defines these phases with three parameters: the learning rate, decay rate, and minimum achievable error rate. This inverse-power-law 'learning' behavior appears to be widespread and has been observed in diverse prediction contexts [48]. It has been utilized for sample size determination in sample classification with microarray data, initially in an unweighted manner by (Mukherjee et al., 2003)[23] and subsequently refined by (Figueroa et al., 2012) [24] using a weighted strategy to favor larger sample sizes in model fitting. Here, we employ the method introduced by (Figueroa et al., 2012) [24] in conjunction with our data augmentation algorithm to relate a learning technique's accuracy to the sample size of transcriptomic sequencing data that is synthesized via data augmentation.

In this article, we present a comprehensive workflow that leverages the SyNG-BTS algorithm and evaluate its performance in augmenting both miRNA-seq and RNA-seq data. Our investigation delves into critical nuances in algorithm specifications, including model choice, hyperparameter tuning, and offline augmentation, alongside key pilot data characteristics, such as sample size, marker filtering, and data normalization. Performance evaluation is grounded in pilot data sourced from the Cancer Genome Atlas (TCGA), which also serves as reference data for comparison with the augmented data [49-55]. We utilize a spectrum of evaluation metrics, including marker-specific summaries, sample clustering, between-marker correlations, and differential expression analysis [56-60]. Furthermore, we extend this workflow to integrate the SyNG-BTS algorithm and the SyntheSize algorithm and assess its performance in *post-hoc* sample size calculation for TCGA studies. We apply this extended workflow to calculate the sample size for developing predictors of immunotherapy response in a study of advanced clear cell renal carcinoma [61], providing insights into study design using machine learning for immunotherapy outcome prediction. This novel workflow effectively bridges a methodological gap, addressing a critical challenge in the design of transcriptomic sequencing studies using machine learning. Their deployment promises to significantly enhance the likelihood of deriving valuable outcome predictors for personalized treatment of patients.

## Results

**Overview of SyNG-BTS**

The objective of SyNG-BTS is to train DGMs on a pilot set of bulk transcriptomic sequencing data and subsequently generate data for any number of samples using the trained model (Fig. 1A). Algorithmically, the training of SyNG-BTS involves two main steps, with the first being optional depending on the pilot data characteristics: (1) offline data augmentation using either an AE head or a Gaussian head; (2) online data augmentation using VAEs, GANs, or Flow-based generative models. We also explored their variants, such as Conditional VAE (CVAE), Wasserstein GAN (WGAN), WGAN with Gradient Penalty (WGANGP), Masked Autoregressive Flow (MAF), Generative Flow with Invertible 1x1 Convolutions (GLOW), and Real-valued Non-Volume Preserving (RealNVP) [29,30,62-67]. For all models and their variants, we evaluated values for two shared hyperparameters: the number of learning epochs and the size of learning batches, along with an additional hyperparameter specific to VAE and CVAE [68,69]. For detailed information on the models and their parameter tuning procedures, please refer to Methods.

We evaluated the performance of SyNG-BTS using data from four TCGA datasets studying Skin Cutaneous Melanoma (SKCM), Acute Myeloid Leukemia (LAML), Breast Invasive Carcinoma (BRCA), and Prostate Adenocarcinoma (PRAD). These datasets served a dual purpose: (1) acting as sources for subsampling to generate pilot datasets, and (2) serving as reference datasets for assessing the augmented data quality (Fig. 1B). To determine the quality of the augmented data, we assessed its congruence with the reference empirical data using five key metrics. These metrics, detailed in Methods, collectively captured marker-specific, inter-marker, and inter-sample data characteristics in both one-group and two-group settings. For miRNA-seq data augmentation, we examined all four TCGA datasets in the one-group setting, presenting the results for SKCM in the main text and that for the other three in the Extended Data File. In the two-group setting, we used the combination of SKCM and LAML datasets (referred to as SKCM/LAML) and the combination of BRCA and PRAD datasets (referred to as BRCA/PRAD), with the results for the latter presented in Extended Data. For RNA-seq data augmentation, which needs larger pilot data than miRNA-seq due to the considerably greater number of markers, we focused on the BRCA and PRAD RNA-seq datasets for the one-group setting (with the PRAD results presented in Extended Data) and the BRCA/PRAD combination for the two-group setting. The process for downloading and preprocessing the TCGA data is outlined in Methods.

**SyNG-BTS successfully augmented one-group miRNA data**

We conducted a comprehensive evaluation of various facets of the augmented data, encompassing marker-specific summary statistics (mean, variation, and sparsity), inter-marker relationships (particularly partial correlation among miRNAs belonging to the same polycistronic clusters), and inter-sample relationships (assessed by clustering the augmented data from SyNG-BTS with the empirical data from TCGA), across the four TCGA studies (Fig. 2 and Extended Data Fig. S1-S4 ). In general, the augmented data exhibited high comparability with the empirical data when suitable DGMs and reasonable pilot data sample sizes were utilized, with the latter depending on the specific DGM. The degree of comparability was further influenced by the interplay of pilot data characteristics and algorithm configurations. Detailed results are presented below in the context of the TCGA SKCM study (Fig. 2 and Extended Data Fig. S1), with similar observations noted in the other three TCGA studies (Extended Data Fig. S2-S4).

Model choice played a crucial role in the augmented data quality (Fig. 2). Among the DGMs examined, VAE (specifically with the ratio between reconstruction loss and Kullback-Leibler divergence being 1:10, shorthanded as VAE1-10) and Flow-models (especially MAF) emerged as the top performers across all evaluation metrics (Fig. 2A-2D). Compared with VAEs, MAF better preserved the proportions of expressed markers (that is, markers with non-zero reads in at least one sample) (Fig. 2C-2D). Furthermore, VAE1-10 excelled in scenarios favoring deep training, typically with a fixed number of epochs or a small batch size, while MAF showed relative insensitivity to batch size and performed well with early stopping (Fig. 2E-2F). Among the GAN-based models, WGANGP outperformed GAN and WGAN across most of the evaluation metrics especially for marker specific mean and sample clustering (Figure 2A-2D). It also showed overall insensitivity to batch sizes and epoch strategies although occasionally favored early stopping (Fig. 2E-2F).

The pilot data characteristic with the most significant impact was sample size (Fig. 2). Increasing the pilot data sample size considerably improved data congruence for VAEs and Flow-based models, as evidenced by enhancements across the evaluation metrics (Fig. 2A-2D). Take VAE1-10 as an example, as the pilot data sample size increased from 20 to 100, the similarity of marker-specific means and standard deviations greatly improved, nearly halving the median absolute deviation between the augmented data and empirical data (Fig. 2A-2B); the mixing of the two data sources upon clustering sharply enhanced, raising the complimentary Adjusted Rand Index (cARI) from about 0.55 to nearly 1; the concordance of inter-marker correlations gradually increased, with the correlation coefficient rising from 0.73 to 0.80 (Fig. 2C). On the other hand, the GAN family performed poorly across all pilot data sample sizes, especially in terms of marker-specific summary statistics (Fig. 2A-2B). Hence, a reasonable sample size (40 or more for VAE1-10 and 60 or more for MAF) proved effective for model training, a phenomenon particularly pronounced for high-performing models like VAEs and Flow-based models.

In addition to the pilot data sample size, we evaluated the impact of marker filtering (Fig. 2B *versus* Fig. 2A, Fig. 2D *versus* Fig. 2C, and Fig. 2E-2H *versus* Fig. S1) and sequencing depth normalization (Fig. 2G and Fig. S1C) for pilot data on the efficacy of model training. The effectiveness of marker filtering (that is, removing markers with consistently low expression across samples) was evident, leading to a substantial enhancements in both the non-zero marker proportions (with its difference between the augmented data and empirical data decreasing from about 25% to 5% for VAE1-10 and from about 15% to 0% for MAF) and the mixing of samples from the two data sources (with the cARI increasing from about 0.55 to 0.95 for VAE1-10 using 40 pilot samples and from around 0.92 to 0.99 for MAF using 60 pilot samples). Its impact on the inter-marker correlation metric varied depending on the model, with notable improvements for Flow-based models. The use of depth normalization may or may not improve data congruence in the one-group setting. Although Trimmed Mean of M-values (TMM) and Total Count normalization outperformed Upper Quartile normalization, they were found to be roughly equivalent or slightly inferior to no normalization.

We further evaluated the impact of offline augmentation on model training (Fig. 2H). Offline augmentation via AE reconstruction proved effective in facilitating the training process, resulting in further enhancement even for the top-performing models like VAE1-10 and MAF. However, it did not improve the performance of GAN models, underscoring the challenges of training GANs in this context.

**SyNG-BTS successfully augmented two-group miRNA data**

For the two-group setting, we replaced VAEs with CVAEs and excluded the GAN models due to their poor performance in the one-group setting. Detailed results are presented below for the TCGA SKCM/LAML study with marker filtering (Fig. 3 and Fig. S5). Similar observations were noted for this study without marker filtering (Fig. S6) and for the BRCA/PRAD study (Fig. S7).

The performance of SyNG-BTS remained consistently strong in the two-group setting (Fig. 3 and Fig. S5). Specifically, MAF once again emerged as the top performer, closely followed by CVAE1-10 (Fig. 3A-3B and Fig S5A). Their performance was influenced by pilot data sample size (Fig. 3A-3B and Fig S5A), marker filtering (Fig. 3 and Fig. S5 *versus* Fig. S6), offline augmentation (Fig. S5B), and hyperparameter tuning (Fig. S5C-S5D), similar to the one-group setting. Additionally, MAF consistently yielded superior results in terms of differential expression analysis, as indicated by the concordance correlation coefficients of p-values (Fig. 3A third row) and group mean differences (Fig. 3A fourth row). Notably, depth normalization, particularly with Total Count or TMM, proved to be more influential than in the one-group setting (Fig. 3B). It played a noticeable role in facilitating model training, especially for CVAE1-10, particularly for the inter-marker correlation metric and the two metrics related to differential expression analysis. The Uniform Manifold Approximation and Projection (UMAP) plot further affirmed the quality of the generated samples, displaying distinct separation by sample types without differentiation according to data sources, even with the runner-up generative model CVAE1-10 (Fig. 3C).

**SyNG-BTS successfully augmented RNA data**

For RNA-seq data augmentation, we focused on the better performing model variant for each DGM model based on the miRNA results, namely VAE, MAF, and WGANGP. Considering the substantial number of markers (60,660) in RNA-seq data, we adjusted the loss ratio of VAE and CVAE to 1:100 (shorthanded as VAE1-100 and CVAE1-100, respectively) and expanded the range of pilot data sample sizes to 50 to 250. Moreover, we excluded markers with both low mean and low variability across samples, reducing the number of markers to 1,099 for the TCGA BRCA data and 1,279 for the BRCA/PRAD data (see details in Methods Table 1).

In the one-group setting, the performance of SyNG-BTS for RNA-seq aligned well with that for miRNA-seq (Fig. 4A-4C and Fig. S8). MAF performed the best for both marker-specific characteristics (Fig. 4A) and sample clustering (Fig. 4B), closely followed by VAE1-100. Like miRNA-seq, depth normalization had little impact on RNA-seq data augmentation in the one-group setting (Fig. 4C).

In the two-group setting, the performance of SyNG-BTS was again consistent with that for miRNA-seq (Fig. 4D-4G and Fig. S9). MAF initially exhibited inferior performance to CVAE when the pilot data sample size was 50 but significantly improved as the sample size increased towards 250 (Fig. 4D-4E). In particular, when the pilot data size exceeded 50, MAF demonstrated exceptional effectiveness in identifying differentially expressed markers between two sample types, achieving nearly perfect agreement with the empirical data in terms of the p-values and fold-changes (Fig. 4E). Both models were highly effective in sample clustering, with the identified clusters showing strong alignment with sample groups rather than data sources (Fig. 4E-4G). The impact of depth normalization is mixed, with Total Count and TMM facilitating smoother improvement over pilot data sample size for MAF (Fig. 4F).

For offline augmentation, the AE reconstruction approach faced challenges due to its complexity and the need for a relatively moderate marker-to-sample-size ratio in the pilot data (results not shown), while Gaussian noise addition proved to be more effective (Fig. 4B and 4E). We used the latter for RNA-seq data offline augmentation by combining an initial pilot dataset with nine noise-added datasets created by introducing Gaussian noise (see details in Methods). This approach reduced variability in all evaluation metrics, thereby improving the quality of the augmented data, especially in the two-group setting (Fig. 4E). Unsurprisingly, the influence of offline augmentation was particularly marked when dealing with small pilot data sizes, with MAF reaping significant benefits in such instances.

**Transfer learning enhanced the performance of SyNG-BTS**

To examine the potential of transfer learning as a pre-training strategy for improving the performance of generative models, we pre-trained VAEs with a loss ratio of 1:10 for miRNA-seq and 1:100 for RNA-seq, using datasets from one TCGA study or multiple studies combined (called the pre-training dataset) [70,71]. The trained models were then used for model training with pilot datasets drawn from a different and intended TCGA study. As shown in Fig. 5, model training saw enhancement across all evaluated pilot data sizes. For miRNA-seq, the enhancement was particularly evident in the improvement of inter-marker relationship (Fig. 5A third row). Conversely, for RNA-seq, the enhancement was more remarkable in preserving the proportion of expressed markers (Fig. 5B first row). While technically any dataset with the same set of markers can be used for pre-training, our findings highlighted the importance of the pre-training data having characteristics comparable to the pilot data (Fig. 5A left column). Additionally, a larger pre-training dataset, such as the combination of TCGA PRAD, LAML, and SKCM data, led to greater enhancements compared to using the TCGA PRAD data alone, when augmenting pilot datasets drawn from the TCGA BRCA study (Fig. 5A right column). These results underscored the value of incorporating transfer learning in transcriptomic data augmentation to leverage distributionally comparable and well-sized pre-training data.

**Overview of SyntheSize**

Our proposed approach for sample size determination using augmented datasets is implemented in four main steps (Fig. 1C).

I. **Data augmentation**: Select a set of candidate sample sizes that are evenly distributed (denoted as $n_i$ for $i = 1, ..., m$) and generate data for each $n_i$ sample size using SyNG-BTS.

II. **Classifier training**: Use each augmented dataset to train a classifier with a chosen learning technique (such as Support Vector Machine) and assess its accuracy. Steps I and II can be repeated for multiple augmented datasets of each sample size $n_i$ to obtain multiple accuracy estimates, providing a more stable average estimate.

III. **Learning curve fitting**: Fit the estimated accuracies for all candidate sample sizes to a learning curve using the IPLF. Its parameters are estimated via a nonlinear weighted least squares optimization, employing the '*nl2sol*' routine from the Port Library, as outlined by (Figueroa et al., 2012) [24]. In this optimization, the weight for the i-th sample size is $i/m$, placing greater emphasis on larger sizes.

IV. **Sample size projection**: Utilizing the fitted curve, the prediction accuracy is projected for any desired sample size, which applies the IPLF with the estimated parameters. In particular, the fitted curve can be used to extrapolate the accuracy level for a larger sample size than $n_m$.

**SyntheSize successfully determined the sample size for miRNA studies**

For demonstration purposes, we applied the SyntheSize approach for (*post-hoc*) sample size evaluation using the TCGA BRCA miRNA-seq data, which includes two subtypes: Invasive Ductal Carcinoma (IDC) and Invasive Lobular Carcinoma (ILC) (Fig. 1D and Fig. 6A). A subset of the TCGA BRCA miRNA-seq data (100 samples per subtype) was reserved as an independent validation set, utilized to provide a *de facto* assessment of the relationship between prediction accuracy and sample size. The remaining samples were then used as the input pilot data for SyNG-BTS as part of the SyntheSize algorithm. We computed accuracies for three learning techniques – Support Vector Machine, K-Nearest Neighbors, and XGBoost – in classifying the two BRCA subtypes. The estimated accuracies fitted well with an IPLF curve, which began to plateau when the sample size reached about 50 per subtype, suggesting limited value in adding more samples (Fig. 6A right column). Additionally, we obtained datasets with varying sample sizes by subsampling the validation set (up to 100 samples per subtype), assessed classification accuracies in these datasets, and fitted IPLF curves for the same three learning techniques (Fig. 6A left column). The curves fitted to the empirical datasets closely mirrored those based on the augmented datasets, providing additional validation for the effectiveness of our proposed approach for sample size determination.

**SyntheSize successfully determined the sample size for RNA studies**

Subsequently, we assessed SyntheSize using the TCGA BRCA RNA-seq data, similar to the assessment with the miRNA-seq data (Fig. 6B). Although exhibiting slightly inferior performance compared to its efficacy for miRNA-seq, SyntheSize provided a satisfactory sample size estimation for RNA-seq. This is evident from the proximity observed between the predicted accuracies using the augmented data and that derived from the empirical validation data.

For further illustration, we showcased SyntheSize in determining the sample size needed for building a predictor of immunotherapy response (Complete/Partial Response *versus* Progressive/Stable Disease), sourcing pilot RNA-seq data from a recent clinical study involving a PD-1 inhibitor, nivolumab, in patients with advanced clear cell renal cell carcinoma[61]. The real and generated samples had a high degree of similarity as revealed in the UMAP (Fig. S10). The accuracies of the three learning techniques again closely aligned with the IPLF model (Fig. 6C). The curves plateaued, indicating that their near-optimal accuracies were achieved, when the sample size reached about 200 samples per response group. Among the three techniques, K-Nearest Neighbors exhibited a noticeably smoother fit to the IPLF curve, albeit with a much higher sensitivity to the sample size as its performance floundered at low sample sizes, compared to Support Vector Machine and XGBoost (Fig. 6C).

## Discussion

Our proposed SyntheSize approach adeptly estimates the required sample size for machine learning with bulk transcriptomic sequencing data, harnessing the power of deep generative models via the SyNG-BTS algorithm to augment available pilot data. The consistent and reliable performance of SyntheSize, demonstrated in both miRNA-seq and RNA-seq, highlights its versatility and effectiveness in informing experimental design for transcriptomics studies using machine learning.

We acknowledge that obtaining pilot data of reasonable size and good quality can be challenging, but it is necessary to avoid making parametric distribution assumptions or relying on a substantial empirical dataset for subsampling. Ideally, users should source their own pilot data that mirrors real-world data characteristics in the intended biomedical problem context. For the large dataset to be collected, for which the sample size is assessed, obtaining a pilot dataset of 40 to 60 samples is worthwhile. If this is not possible, users can turn to publicly available data. For instance, the TCGA offers high-quality transcriptomic sequencing data for more than 30 cancer types, each with hundreds of samples.

Through a comprehensive evaluation of SyNG-BTS in diverse settings, we have demonstrated the successful training of generative models for bulk transcriptomic sequencing data. The efficacy of these models is influenced by various factors related to the pilot data, such as its sample size and marker numbers, as well as specifications for the generative models, including model choice, hyperparameter tuning, and the use of offline augmentation and transfer learning. Generally, model training is more successful when the pilot data maintains a reasonable marker-to-sample-size ratio. In cases where this ratio is excessively high, the incorporation of offline augmentation and transfer learning has proven to be beneficial. The generative models need to be thoughtfully selected and meticulously tuned. Among the models investigated, MAF and VAE models consistently outperformed GAN models.

The runtime of the DGMs used in SyNG-BTS is an important consideration for its overall utility. In practice, the time required to train these models can vary based on data complexity, model architecture, and available computational resources. Our experiences found that the DGMs did not demand extensive computational resources, primarily due to the simplicity of the model structures employed and the modest size of the pilot datasets involved. Specifically, when using any of the DGMs in our studies, the runtime typically ranges between 1 to 5 minutes on a personal computer with 16GB of RAM and a 2.3 GHz Quad-Core Intel Core i5 processor. This brief runtime indicates the manageability of these models, affirming that even personal computers, without parallel computing setups, are sufficient for training and applying the DGMs. The low computational demands significantly broaden the potential for using SyntheSize to design transcriptomic sequencing studies, without necessitating high-end computing infrastructure.

In summary, our study demonstrated the successful training of generative models to augment bulk-tissue transcriptomic sequencing data, enabling effective sample size determination using augmented datasets and the IPLF model. These computational resources are poised to greatly facilitate the deployment of supervised machine learning techniques in deriving effective sample classifiers from biomedical transcriptomic data. These contributions will significantly advance the development of essential computational tools crucial for designing classification studies with transcriptomic sequencing data, thereby accelerating their translation into clinically impactful predictors.

## Methods

**SyNG-BTS**

We introduced the SyNG-BTS (Synthesize Next Generation Bulk Transcriptomic Sequencing) algorithm to augment transcriptomic sequencing data from a pilot dataset using deep generative models like Variational Autoencoder (VAE), Generative Adversarial Network (GAN), and Flow-based models. For each model, we explored different variants and fine-tuned hyperparameters. Additionally, we investigated the effectiveness of utilizing offline augmentation and transfer learning to enhance model training, especially when the pilot data has a low sample size relative to the number of markers. The neural net structures of the three generative models are presented in Figure S11-13. They employed the Adam optimizer and ReLU activation function, with the learning rate fixed at 0.0005.

**Tuning of shared hyper-parameters**. The three generative models have two shared hyper-parameters – the number of learning epochs and the size of learning batches. Both wield significant influence on the training depth and require careful fine-tuning. Small batches with fixed epochs have a similar effect to more learning epochs, so it is beneficial to explore complex epoch settings rather than rely on a fixed number of epochs. In addition to fixed learning epochs (200 for Flow-based models and 1000 for VAEs and GANs), we considered an early stopping rule, adopting the method proposed in (Li et al., 2021) [72]– if the loss does not improve for 30 epochs, then stop. For the learning batch size, we examined candidate batches at 10% and 20% of the training data sample size.

**Model variants and tuning of variant-specific hypermeters.** We explored the following variants for each generative model utilized in the SyNG-BTS algorithm.

- **VAE**: VAE is a generative model that learns to encode and decode data by mapping it to a probabilistic latent space. Its loss evaluation consists of two main components: a reconstruction loss, measuring how well the model reconstructs the input data, and a Kullback-Leibler (KL) divergence term, enforcing the learned latent space to follow a predefined probability distribution. The balance between the reconstruction loss and the KL divergence is adjustable, allowing us to prioritize the faithfulness of reconstructions or the disentanglement of latent variables. In SyNG-BTS, the ratio of reconstruction loss to KL divergence is set to 1:1, 1:5, and 1:10 for microRNA-seq (miRNA-seq) data and 1:100 for RNA-seq data. The corresponding VAE models are named VAE1-1, VAE1-5, VAE1-10, and VAE1-100, respectively. Conditional VAE (CVAE) extends the VAE by introducing conditional information during both the encoding and decoding processes. This enables the generation of data conditioned on specific attributes or categories, making it suitable for generating data with multiple sample groups.

- **GAN**: GAN is a neural network architecture comprised of a generator and a discriminator engaged in a competitive-game-like setting. The generator aims to produce data that are indistinguishable from real-world data, while the discriminator works to differentiate between real and generated data. Two GAN variants explored in this study are Wasserstein GAN (WGAN) and WGAN with Gradient Penalty (WGANGP). WGAN retains the basic GAN structure but employs Wasserstein distance as a more stable and meaningful measure of the difference between generated and real data distributions, addressing training instability issues. WGANGP enhances WGAN by imposing a Lipschitz continuity constraint through a gradient penalty term in the loss function, further improving training stability. In the regular GAN model, each iteration of the generator is followed by one iteration of the discriminator, while WGAN and WGANGP set this iteration ratio to 5, following the original papers [63,64]. In our study, the weight clipping parameter for WGAN is set to 0.01, and the gradient penalty parameter $\lambda$ for WGANGP is set to 10.

- **Flow-based models**: Flow-based models utilize invertible mappings to transform complex data distributions into simpler, tractable latent spaces. This paper explores three such models: Real-Valued Non-Volume Preserving (RealNVP), Generative Flow with Invertible 1x1 Convolutions (GLOW), and Masked Autoregressive Flow (MAF). While RealNVP and GLOW share a similar network structure, GLOW enhances it with '1x1 convolutions' and 'actnorm' layers for improved performance, albeit requiring larger sample sizes due to its deeper structure. MAF, distinct from RealNVP and GLOW, leverages the autoregressive property by conditioning each variable on its preceding neural network layers and applying transformations across all data dimensions, thereby enhancing the model's capacity to capture complex distributions. These Flow-based models prioritize invertibility, enabling efficient sampling and precise likelihood computation. In our study, the number of blocks, representing the sequential layers within the model that process the data, for these models is set to 5, as in the original paper [41]. The validation set ratio, which is the proportion of the dataset reserved for validating the model's performance, is fixed at 0.15.

**Offline augmentation**. We examined two techniques for offline augmentation: (1) applying an autoencoder (AE) head for pilot data with a relatively modest number of markers compared to the number of samples [44], and (2) adding Gaussian white noise for pilot data with a relatively large marker-to-sample-size ratio [46]. The AE head iteratively reconstructs the input data and combine the reconstructed data with the data from the previous iteration, exponentially increasing the number of samples. Assuming the pilot data sample size is k, AE offline augmentation with $t$ iterations would lead to a sample size of $k * 2^t$. The value of $t$ is set to be 2 in our study. Gaussian white noise addition involves adding noise generated from a Gaussian distribution to the pilot sequencing data on the log2 scale, and this process can be repeated multiple times with the resulting datasets combined to increase the sample size. While offline augmentation generates dependent samples, leveraging dependent samples poses no issues in deep neural networks. This is exemplified by common practices in image data augmentation, such as flipping, cutting, or color changing.

**Transfer learning.** Transfer learning is a machine learning approach that boosts model training efficiency by leveraging knowledge acquired from one task to improve performance on another related task [73]. Instead of training models from scratch for a specific objective, transfer learning involves pre-training a model on a large dataset for a source task. The knowledge gained during this pre-training is then fine-tuned for a target task with a (often smaller) training dataset. This approach is particularly valuable when data is scarce for the target task, as it allows models to benefit from previously learned features and representations. Transfer learning has demonstrated notable success in diverse domains, including computer vision, natural language processing, and single-cell sequencing, offering advantages such as reduced data requirements and accelerated training convergence, ultimately contributing to enhanced model performance.

## Performance Evaluation of SyNG-BTS

**Datasets.** To assess the performance of SyNG-BTS for augmenting bulk transcriptomic sequencing data, we utilized miRNA-seq and RNA-seq datasets from four TCGA studies: Breast Invasive Carcinoma (BRCA), Prostate Adenocarcinoma (PRAD), Skin Cutaneous Melanoma (SKCM), and Acute Myeloid Leukemia (LAML). These datasets served a dual purpose: (1) as sources for subsampling to create pilot datasets of varying sample sizes, and (2) as reference datasets for evaluating the quality of the augmented data. Data retrieval from TCGA was conducted using the R package *TCGAbiolinks* [51]. Each dataset contains a total of 1,881 markers for miRNA-seq and 60,660 for RNA-seq. The sample sizes for miRNA-seq in BRCA, PRAD, SKCM, and LAML are 1207, 551, 452, and 188, respectively. For RNA-seq, the sample sizes are 1231, 554, 473, and 151, respectively. In the one-sample-group setting, all four datasets were utilized for miRNA-seq and only the BRCA and PRAD datasets were used for RNA-seq given their larger sample sizes. In the two-group setting, we considered the combination of SKCM and LAML (referred to as SKCM/LAML) and the combination of BRCA and PRAD (referred to as BRCA/PRAD) for miRNA-seq and focused on the BRCA/PRAD combination for RNA-seq. Among the miRNA-seq datasets, the single-group evaluation was comprehensively performed for SKCM and BRCA to identify a preferred model training setting. This setting was subsequently used for evaluating LAML and PRAD. Similarly, the two-group evaluation was done comprehensively for SKCM/LAML and selectively for BRCA/PRAD.

**Depth normalization.** Depth normalization is an important step for preprocessing miRNA and RNA sequencing data [53]. To assess the impact of depth normalization on generative model training, we applied the following depth normalization methods to the pilot data: Total Count (TC) [74], Trimmed Mean of Mvalues (TMM) [75], and Upper Quartile (UQ) [52]. Subsequently, the normalized data underwent log2 transformation, with a pseudo count of one added to all counts to address the zero-count issue. Additionally, we refrained from standardizing the input data for generative models to ensure that the generated samples can be transformed back to the original sequencing counts as required by many downstream analyses including differential expression analysis.

**Marker filtering.** MiRNA-seq and RNA-seq data often include a large number of markers with consistently low abundance across samples. Such poorly expressed markers pose a challenge for generative model training while providing minimal value for downstream analyses. To address this, we investigated filtering out poorly expressed markers to improve the performance of SyNG-BTS. For miRNA-seq, where mean and standard deviation exhibit a strong association [76], we used marker-specific means to identify poorly expressed markers, setting a threshold based on mean count. In the case of RNA-seq, both marker-specific means and standard deviations were considered for identifying poorly expressed markers. Our goal was to choose thresholds that retain at least 256 markers, corresponding to the largest number of neurons in the neural network layers. The specific thresholds varied depending on the datasets. Table 1 lists the thresholds used for the four TCGA datasets and another dataset from an immunotherapy dataset [61] that was collected in clear cell renal cell carcinoma (CCRCC) and used for our method application, along with the numbers of remaining markers after filtering.

**Table 1**: Threshold (on the log2 transformed counts) for marker filtering and number of remaining markers.

| Molecule type | Cancer type | Mean threshold | SD threshold | Number of remaining markers |
|---|---|---|---|---|
| miRNA-seq | SKCM | 4 | | 298 |
| | LAML | 2 | | 267 |
| | BRCA | 3 | | 289 |
| | PRAD | 3 | | 268 |
| | SKCM/LAML | 3 | | 317 |
| | BRCA/PRAD | 3 | | 279 |
| RNA-seq | BRCA | 5 | 2 | 1099 |
| | PRAD | 5 | 2 | 460 |
| | BRCA/PRAD | 5 | 2 | 1279 |
| | CCRCC (CRPR/PDSD) | 30 | 3 | 259 |

**Transfer learning**. In scenarios where the pilot data has a relatively high marker number to sample size ratio, we explored the use of transfer learning to enhance generative model training by leveraging pre-trained models. For miRNA-seq, we considered four distinct scenarios: (1) training on SKCM pilot datasets with transfer from the LAML dataset, (2) training on SKCM pilot datasets with transfer from the combined LAML/BRCA/PRAD dataset, (3) training on BRCA pilot datasets with transfer from the PRAD dataset, and (4) training on BRCA pilot datasets with transfer from the LAML/SKCM/PRAD dataset. The key differences between the two pre-training datasets lie in the trade-off between sample size and the level of data heterogeneity. For RNA-seq, we trained BRCA pilot datasets with transfer from the PRAD dataset and *vice versa*. Model pre-training utilized the VAE models, with a batch size set at 10% of the pre-training data size, an epoch at 1000, and a learning rate of 0.0005.

**Data augmentation.** To generate pilot datasets, we randomly draw five pilot datasets with a pre-specified sample size from an empirical dataset. Each pilot dataset was utilized to train a generative model with specified values of epochs, learning rate, and learning batch size. Subsequently, each trained model was used to create new samples at a pre-defined target sample size. For each TCGA dataset, pilot data size, generative model choice, and hyper-parameter setting, we created 25 sets of augmented data, each matching the sample size of the selected dataset from which the pilot dataset was drawn. The pilot data sample sizes per group were examined at (1) 20, 40, 60, 80, and 100 for miRNA-seq, and (2) 50, 150, and 250 for RNA-seq.

**Evaluation of augmented data.** We evaluated the quality of the augmented data by comparing them with the TCGA empirical data using five key analysis metrics. These metrics collectively captured marker-specific, inter-marker, and inter-sample data characteristics, in both one-group and two-group settings.

- *Marker-specific summary*. We compared marker-specific summary statistics, including (1) mean, (2) standard deviation, (3) sparsity, defined as the percentage of zero counts across all samples, between the generated and empirical data. The preservation of marker-specific characteristics was gauged with the Median Absolute Deviation (MAD) between the two sets of summary statistics [36,56]. A MAD close to zero indicates a high degree of data comparability. In addition, we compared the percentage of markers that have zero counts across all samples between the data sources.
- *Inter-marker correlation*. We examined the degree of correlation among well-expressed miRNAs that belong to the same polycistronic clusters, known for their tendency to be co-expressed, as previously reported [57]. We quantified their correlation with the partial correlation coefficient (PCC). We assessed its level of agreement between the generated and empirical data using the concordance correlation coefficient (CCC) [58]. A CCC close to 1 signifies a high level of data comparability.

- *Sample clustering*. We conducted hierarchical clustering [77] on a combined dataset comprising both generated and empirical samples. The clustering used the Euclidean distance measure and the Ward linkage function. To assess the alignment of clusters, we utilized the Adjusted Rand Index (ARI), treating the data sources as ground truth labels in the single-group setting [59,78]. In this context, an ARI close to 0 suggests comparability between the two data sources. For consistence with the favorable direction of other metrics, we converted it to 1 – ARI, referred to as the complementary ARI (cARI). In the two-group setting, we used the ARI treating the sample groups as the ground truth to evaluate the quality of the augmented samples. An ARI close to 1 indicates that the sample separation is predominantly driven by biological differences between the sample groups, rather than technical differences between the two data sources.

- *Dimension reduction*. We performed dimension reduction using the Uniform Manifold Approximation and Projection (UMAP) method [60]. UMAP is extensively employed in single-cell studies to uncover a low-dimensional representation that closely approximates the fuzzy topological structure, aiding in data visualization and pattern diagnosis. UMAP plots were utilized to visually assess the separation of samples by the two data sources.

- *Differential expression analysis*. To assess evidence of differential expression in datasets that include two sample types, we applied the voom method to each data source [79], using the *DE.voom* function in the *PRECISION.seq* package [53,54]. Subsequently, we compared the two sets of results for differential expression analysis, including p-values and fold-changes, using the CCC. A CCC close to 1 indicates a high level of data congruence.

**SyntheSize**

We selected m equally spaced candidate sample sizes. For each sample size $n_i$, where $i = 1, \ldots, m$, we generated $n_i$ new samples with the trained generative model. Subsequently, we utilized the generated samples to build a classifier using machine learning techniques, such as Support Vector Machine, and assessed its classification accuracy. This entire process was repeated 30 times for each candidate sample size, and the average accuracy was computed. To quantify the relationship between accuracy and sample size, we adopted the method introduced in (Figueroa et al., 2012) [24], fitting a learning curve using the inverse power law functions (IPLFs) as follows:

$$accuracy = (1 - a) - b(sample\ size)^c.$$

The parameters a, b, c are estimated through nonlinear weighted least squares optimization using the '*nl2sol*' routine from the Port Library [80], where the weight for the i-th candidate sample size is $i/m, i = 1, \ldots, m$. These weights emphasize that the accuracy is more reliable for larger sample sizes. After fitting the learning curve, we can predict the accuracy for uncalculated sample sizes, along with the 95% prediction interval, following the approach outlined in (Figueroa et al., 2012) [24].

**Performance Evaluation of SyntheSize**

For illustration and evaluation purposes, we applied SyntheSize for *post-hoc* sample size evaluation in the TCGA BRCA study to classify its two subtypes, Invasive Ductal Carcinoma (IDC) *versus* Invasive Lobular Carcinoma (ILC), using miRNA-seq data (n = 871 for IDC and n = 210 for ILC) and RNA-seq data (n = 892 for IDC and n = 213 for ILC). In addition to marker filtering based on marker-specific means and standard deviations, we further selected markers based on the importance score from a random forest analysis to adjust the signal to noise ratio.

For both miRNA-seq and RNA-seq data, we initially reserved 100 IDC and 100 ILC samples as an independent validation set, which would be utilized to estimate the *de facto* IPLF of classification accuracy. The remaining samples were then employed to draw pilot data for SyNG-BTS as part of the SyntheSize algorithm. CVAE1-20 with 285 epochs was employed for miRNA-seq, and CVAE1-50 with 185 epochs was employed for RNA-seq. The batch fraction was set at 10%, and the learning rate was fixed at 0.0005.

We considered three commonly used classifiers: Support Vector Machine (implemented by the R package *e1071*), K-Nearest Neighbors with K=20 (implemented by the R package *class*), and XGBoost with 25 rounds for miRNA data and 10 rounds for RNA data (implemented by the R package *xgboost*). For RNA-seq data, specific XGBoost parameters were adjusted: the learning rate was set to 0.1, the maximum depth of a tree was set to 3, and the minimum sum of instance weight (hessian) needed in a child was set to 3; other training parameters were kept at their default values. The classifiers' performance was assessed based on classification accuracy computed through 5-fold cross-validation, given the limited validation data size.

We trained classifiers using generated samples (across a range of candidate sample sizes) and using real samples (over the same candidate sample size range) from the independent validation set. The accuracies of the classifiers at each candidate sample size were utilized to fit the IPLF. Subsequently, the fitted functions derived from generated samples were compared with those from real samples. This analysis provides insights into the effectiveness and reliability of the SyntheSize algorithm for determining sample size in supervised machine learning with transcriptomic sequencing data.

**Application of SyntheSize to an Immunotherapy Study**

To further illustrate, we utilized SyntheSize for sample size assessment in predicting patient response to a PD-1 inhibitor, nivolumab, with RNA-seq in advanced clear cell renal cell carcinoma. The objective was to build a classifier with RNA-seq data to distinguish two clinical response groups according to RECIST 1.1: Complete or Partial Response (CR/PR) *versus* Stable or Progressive Disease (PD/SD), as outlined in the original paper. Pilot data came from a recent study of advanced clear cell renal cell carcinoma involving 152 patients (39 CR/PRs and 113 PD/SDs), all treated with nivolumab and with available RNA-seq data[61]. For data augmentation, we employed SyNG-BTS using CVAE1-200 with Gaussian head offline augmentation, a batch fraction of 10%, and an early stopping strategy. Subsequently, the augmented data was utilized by SyntheSize to evaluate the classification accuracy using Support Vector Machine, K-Nearest Neighbors, and XGBoost, across a range of total sample sizes from 30 to 400.

## Data Access

The microRNA and RNA sequencing data used in this article are all publicly available, including those generated by the TCGA Research Network: https://www.cancer.gov/tcga.

## Code Availability

The Python and R code for implementing SyNG-BTS and SyntheSize is freely downloadable on GitHub (https://github.com/LXQin/SyNG-BTS and https://github.com/LXQin/SyntheSize). The R code for reproducing the results, which includes TCGA and immunotherapy study data downloading, data augmentation with SyNG-BTS, sample size assessment with SyntheSize, as well as the generation of all figures, can be found at https://github.com/LXQin/SyntheSize-paper-supplementary-materials.

## Competing Interest Statement

The authors declare no competing interests.

## Acknowledgements

This work was supported by grants from the National Institutes of Health: YQ and LXQ were supported by HG012124; XW and LXQ were supported by CA214845; LXQ was also supported by CA008748. We thank Nicole Rusk for her insightful comments on the manuscript and Joseph Kanik for his assistance with the graphical design of Figure 1.

## Author Contributions

**Conceptualization**: YQ and LXQ; **Methodology**: YQ, XW, and LXQ; **Data Curation**: YQ and LXQ; **Formal Analysis**: YQ, XW, and LXQ; **Writing – Original Draft**: YQ and LXQ; **Writing – Review and Editing**: YQ, XW, and LXQ; **Resources**: LXQ; **Supervision**: LXQ; **Funding Acquisition**: LXQ## References

1. van't Veer, L.J. & Bernards, R. Enabling personalized cancer medicine through analysis of gene-expression patterns. *Nature* 452, 564-570 (2008).
2. Adams, J.U. Genetics: Big hopes for big data. *Nature* 527, S108-109 (2015).27

40. Kingma, D.P. & Welling, M. Auto-encoding variational bayes. *arXiv preprint arXiv:1312.6114* (2013).

41. Dinh, L., Krueger, D. & Bengio, Y. Nice: Non-linear independent components estimation. *arXiv preprint arXiv:1410.8516* (2014).

42. Bartel, D.P. MicroRNAs: genomics, biogenesis, mechanism, and function. *Cell* 116, 281-297 (2004).

43. Ambros, V. The functions of animal microRNAs. *Nature* 431, 350-355 (2004).

44. Davila Delgado, J.M. & Oyedele, L. Deep learning with small datasets: using autoencoders to address limited datasets in construction management. *Applied Soft Computing* 112, 107836 (2021).

45. Kramer, M.A. Nonlinear principal component analysis using autoassociative neural networks. *AIChE journal* 37, 233-243 (1991).

46. Bishop, C.M. Training with Noise is Equivalent to Tikhonov Regularization. *Neural Computation* 7, 108-116 (1995).

47. Holmstrom, L. & Koistinen, P. Using additive noise in back-propagation training. *IEEE Trans Neural Netw* 3, 24-38 (1992).

48. Wickens, C.D., Helton, W.S., Hollands, J.G. & Banbury, S. *Engineering psychology and human performance*, (Routledge, 2021).

49. Network, C.G.A.R. Comprehensive genomic characterization defines human glioblastoma genes and core pathways. *Nature* 455, 1061-1068 (2008).

50. Chu, A*., et al.* Large-scale profiling of microRNAs for The Cancer Genome Atlas. *Nucleic Acids Res* 44, e3 (2016).

51. Colaprico, A*., et al.* TCGAbiolinks: an R/Bioconductor package for integrative analysis of TCGA data. *Nucleic Acids Res* 44, e71 (2016).
31

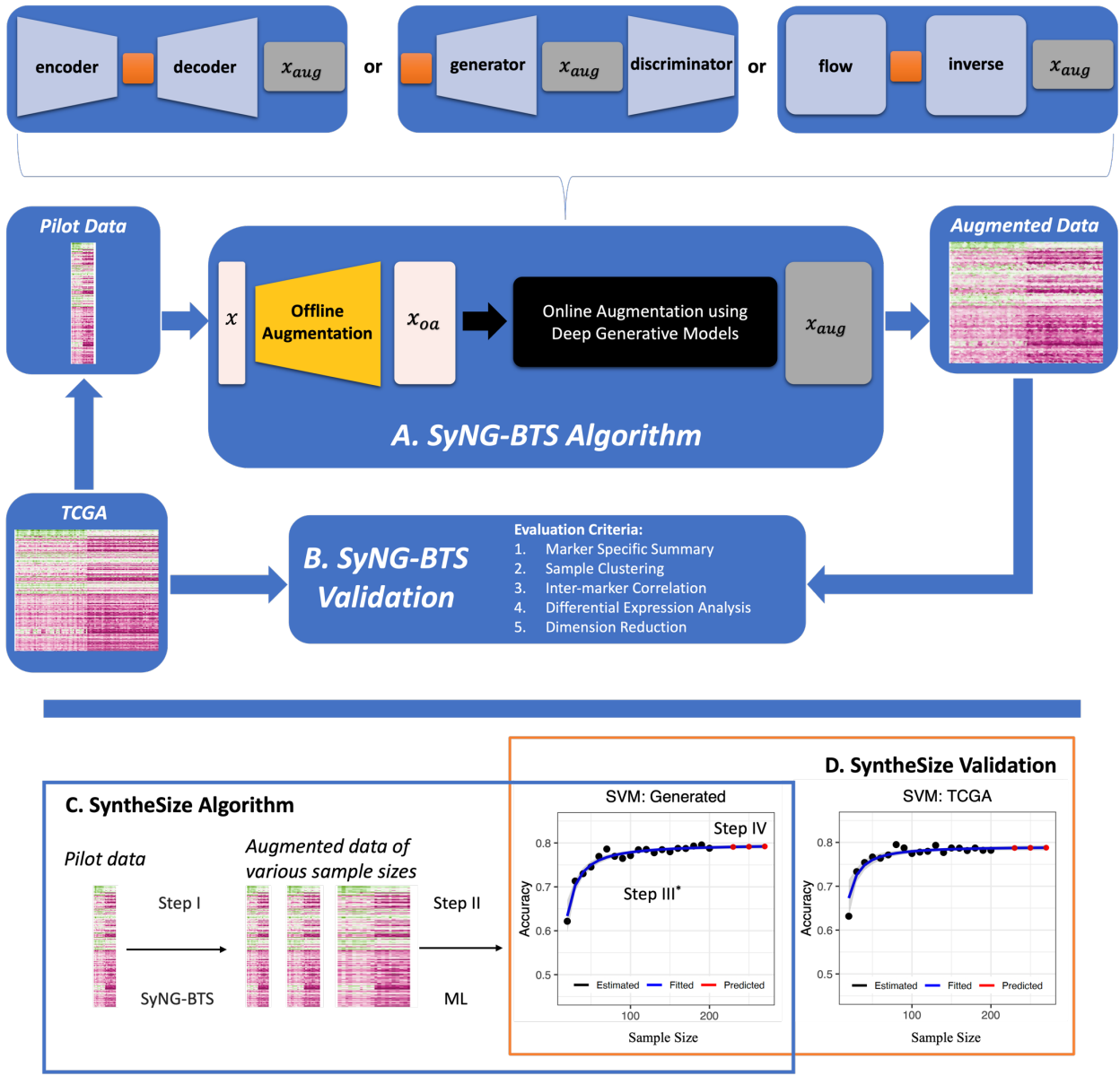

**Fig. 1 | Schema of SyNG-BTS and SyntheSize algorithms along with their respective validation.** **A**: SyNG-BTS algorithm. A pilot dataset, denoted as $x$, undergoes offline augmentation to transform into $x_{oa}$, which subsequently serves as the input for a deep generative model, leading to augmentation to $x_{aug}$. **B**: SyNG-BTS validation. Pilot datasets are generated by randomly sampling from a dataset in the Cancer Genome Atlas (TCGA) and are subsequently inputted into SyNG-BTS to generate augmented data with the matching sample size of the source TCGA dataset, enabling a comprehensive comparison of the empirical data and the augmented data using various evaluation criteria. **C**: SyntheSize algorithm. It determines the sample size for machine learning (ML), such as Support Vector Machine (SVM), in four steps: (I) A pilot dataset is augmented with SyNG-BTS and sampled to generate datasets with a range of sample sizes; (II) Each augmented dataset is used to train a classifier with a chosen ML technique

(such as SVM) and assess its accuracy; (III) The estimated accuracies for various sample sizes are fitted to a learning curve using the Inverse Power Law Function (IPLF); and (IV) Utilizing the fitted curve, the prediction accuracy is projected for any desired sample size. **D**: SyntheSize validation. ML is used to classify the samples in the TCGA breast cancer study to two subtypes, invasive ductal carcinoma and invasive lobular carcinoma. The data is split to two portions: one is used to supply pilot data and derive the IPLF curve with SyntheSize, and the other is to provide empirical datasets of various sample sizes via subsampling to construct another IPLF curve for comparison with the SyntheSize-derived IPLF curve.

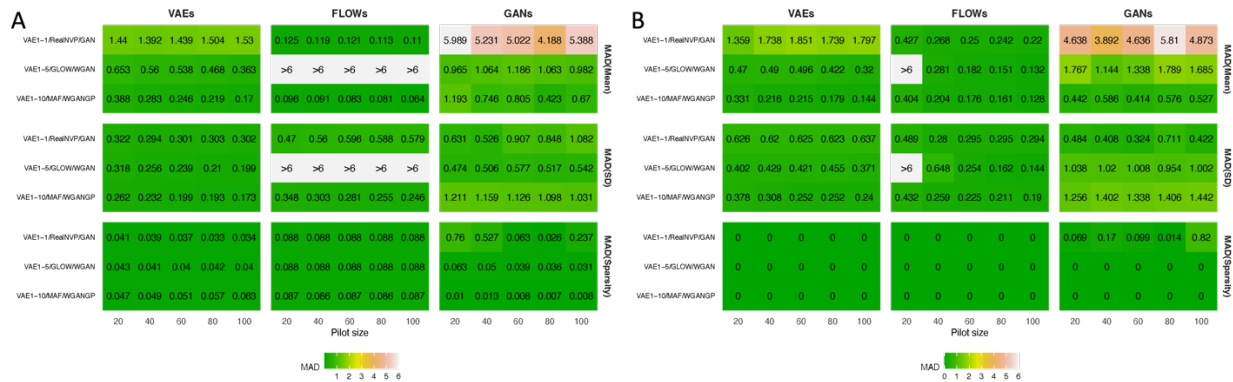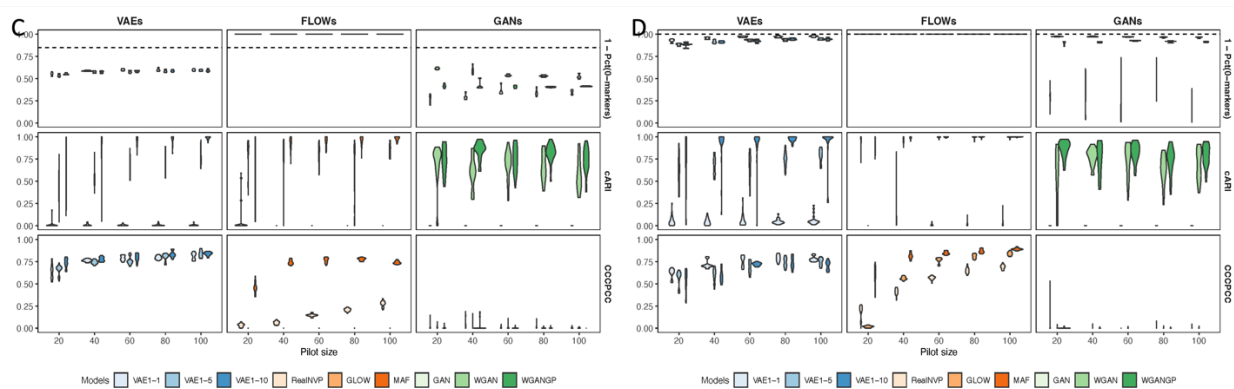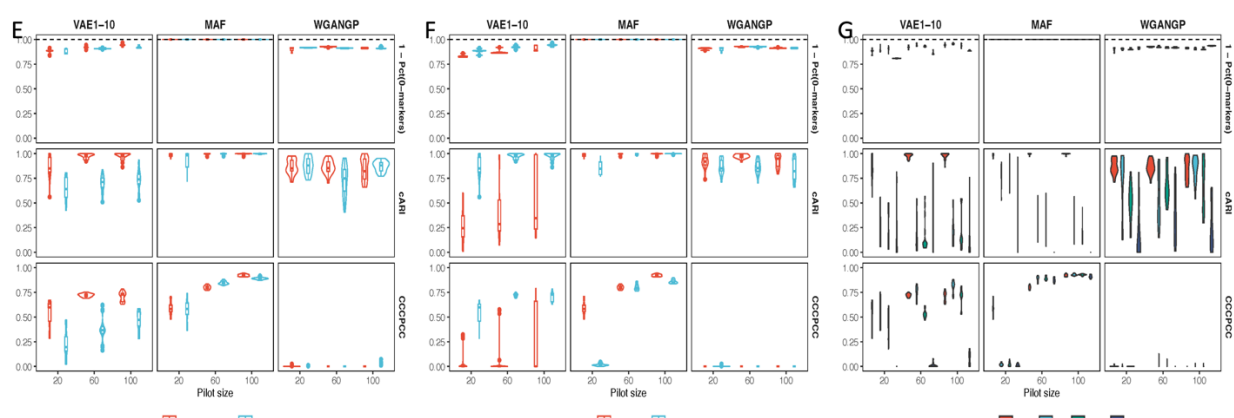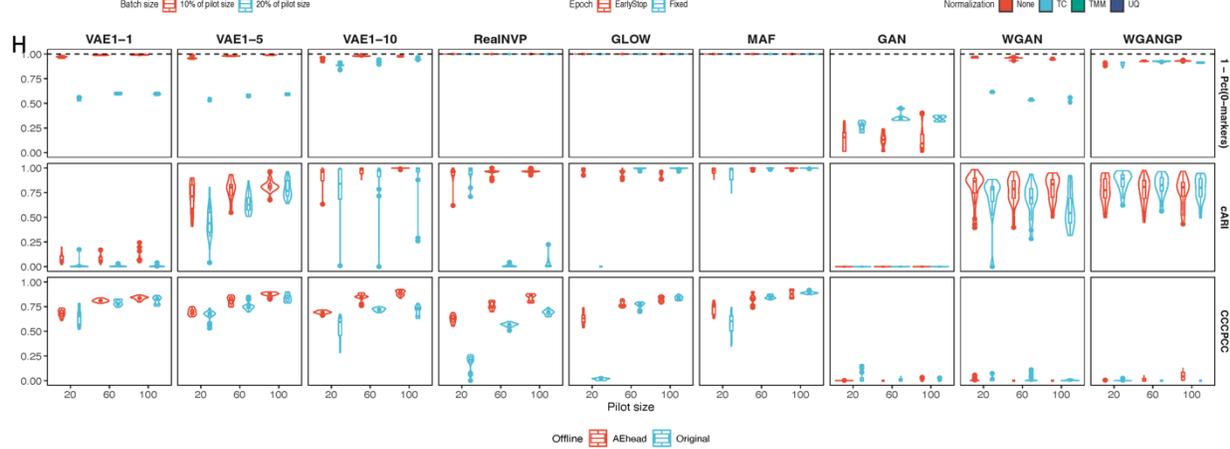

Fig. 2 | SyNG-BTS evaluation for microRNA-seq in the one-group setting, using pilot data from the TCGA SKCM study without marker filtering (Panels A and C) and with marker filtering (Panels B and D-H). **A**: Median Absolute Deviations (MADs) in marker-specific summary statistics (mean, standard deviation, and sparsity, defined as the percentage of zeros) between the SyNG-BTS augmented data and the empirical data are calculated as the pilot data sample size increases from 20 to 100. The MAD values are color-coded, with extremely large values represented as ">6". Smaller MADs indicate better congruency between the augmented data and the empirical data. Each sub-panel column represents one of the three generative model families, and each row within a sub-panel corresponds to a specific model variant, as indicated on the left of each sub-panel. VAE1-10, MAF, and WGANGP consistently exhibit the smallest MAD values in their respective model families. **B**: MADs in marker-specific statistics between the augmented data and the empirical data are evaluated when marker filtering is applied to pilot data. **C**: Additional evaluation metrics, encompassing (1) the percentage of markers with non-zero counts in at least one sample (indicated as 1 – Pct(0-markers)), (2) the agreement of sample clusters and data sources when clustering a combined dataset of both generated and real samples, measured by the complementary Adjusted Rand Index (cARI), and (3) the degree of correlation among member microRNAs belonging to the same polycistronic clusters, quantified by the Concordance Correlation Coefficient of Partial Correlation Coefficients (CCCPCC), are calculated across various pilot data sample sizes. Proximity of values for 1 – Pct(0-markers) to its level in the empirical data (indicated with a horizontal dashed line), along with elevated values of cARI and CCCPCC, signify improved congruency between the augmented data and the empirical data. Flow-based models exhibit smaller non-zero marker proportions than the empirical data, as they are above the dashed line; VAEs tend to generate approximately 50% of markers with zero counts in all samples, while GANs show the highest proportion of zero-count markers. In general, VAE and Flow-based models outperform GAN models, with VAE1-10, MAF, and WGANGP emerging as the top performer in their respective model families. **D**: The same additional evaluation metrics, including 1 – Pct(0-markers), cARI, and CCCPCC, are computed when marker filtering is applied to pilot data. **E**: Evaluation metrics, including 1 – Pct(0-markers), cARI, and CCCPCC, are presented for the best performing variant in each generative model family, using two different training batch size (indicated by colors). VAE1-10 tends to be most sensitive to batch size, showing better performance for smaller batch sizes (that is, deep training), while MAF and WGANGP tend to be insensitive. **F**: Evaluation metrics, including 1 – Pct(0-markers), cARI, and CCCPCC, are presented for the best performing variant in each generative model family, using two different epoch strategies (indicated by colors). VAE1-10 prefers fixed epochs, while MAF already performs well with early stopping. **G**: Evaluation metrics, including 1 – Pct(0-markers), cARI, and CCCPCC, are presented for the best performing variant in each generative model family, using three different depth normalization methods (indicated by colors): Total Count (TC), Trimmed Mean of M-values (TMM), and Upper Quartile (UQ), in comparison with no normalization (None) for pilot data. It is noteworthy that depth normalization has minimum impact on the generative model performance in this context. **H**: Evaluation metrics, including 1 – Pct(0-markers), cARI, and CCCPCC, are presented with or without the use of offline augmentation via AE head (indicated by colors). It is evident that offline augmentation

consistently improves the performance of all three generative models across evaluation metrics in terms of both the average value and variability. Unless stated otherwise, panels **A to G** employ no offline augmentation, no depth normalization, a 10% batch fraction, a fixed epoch strategy for VAEs and GANs, and an early stopping strategy for Flow-based models.

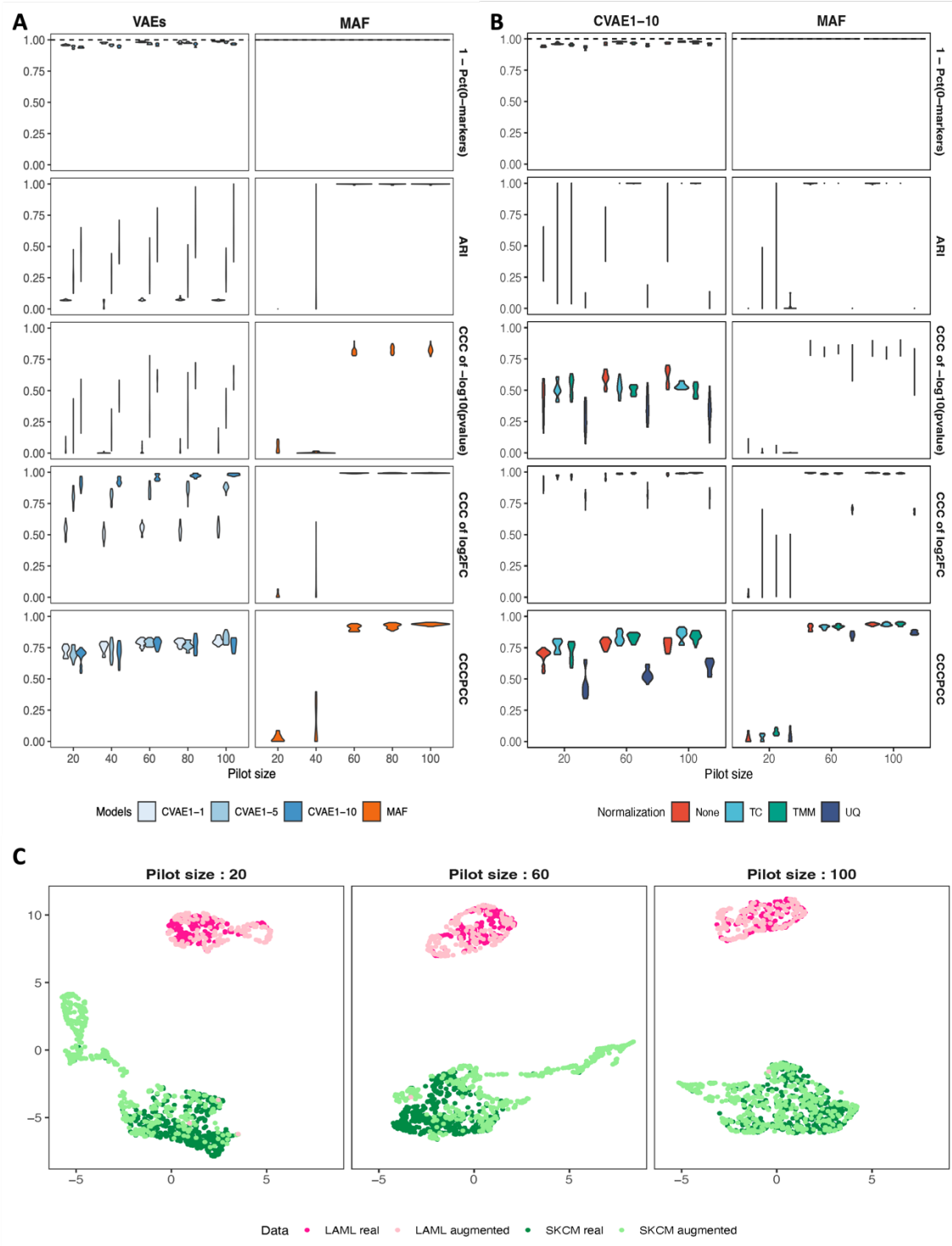

Fig. 3 | SyNG-BTS evaluation for microRNA-seq in the two-group setting, using pilot data from the combination of the TCGA SKCM and LAML studies with marker filtering. A: Evaluation metrics assessing the congruence between the augmented data and the empirical data, including (1) 1 – Pct(0-markers), (2) the agreement of sample clusters and sample types when clustering a combined dataset of both generated and real samples, measured by the Adjusted Rand Index (indicated as ARI), (3) concordant

correlation coefficient of p-values from differential expression analysis on the –log10 scale (indicated as CCC of –log10 (p-value)), (4) concordant correlation coefficient of log2 fold change from differential expression analysis (indicated as CCC of log2FC), and (5) CCCPCC, are calculated for various generative models as the pilot data sample size increases from 20 to 100 per sample group. **B**: The same evaluation metrics for data congruence are calculated using three different depth normalization methods (indicated by colors) in comparison with no normalization. **C**: The Uniform Manifold Approximation and Projection (UMAP) representation for the generated samples (by CVAE1-10) and the real samples, with the data source and the sample type indicated by colors.

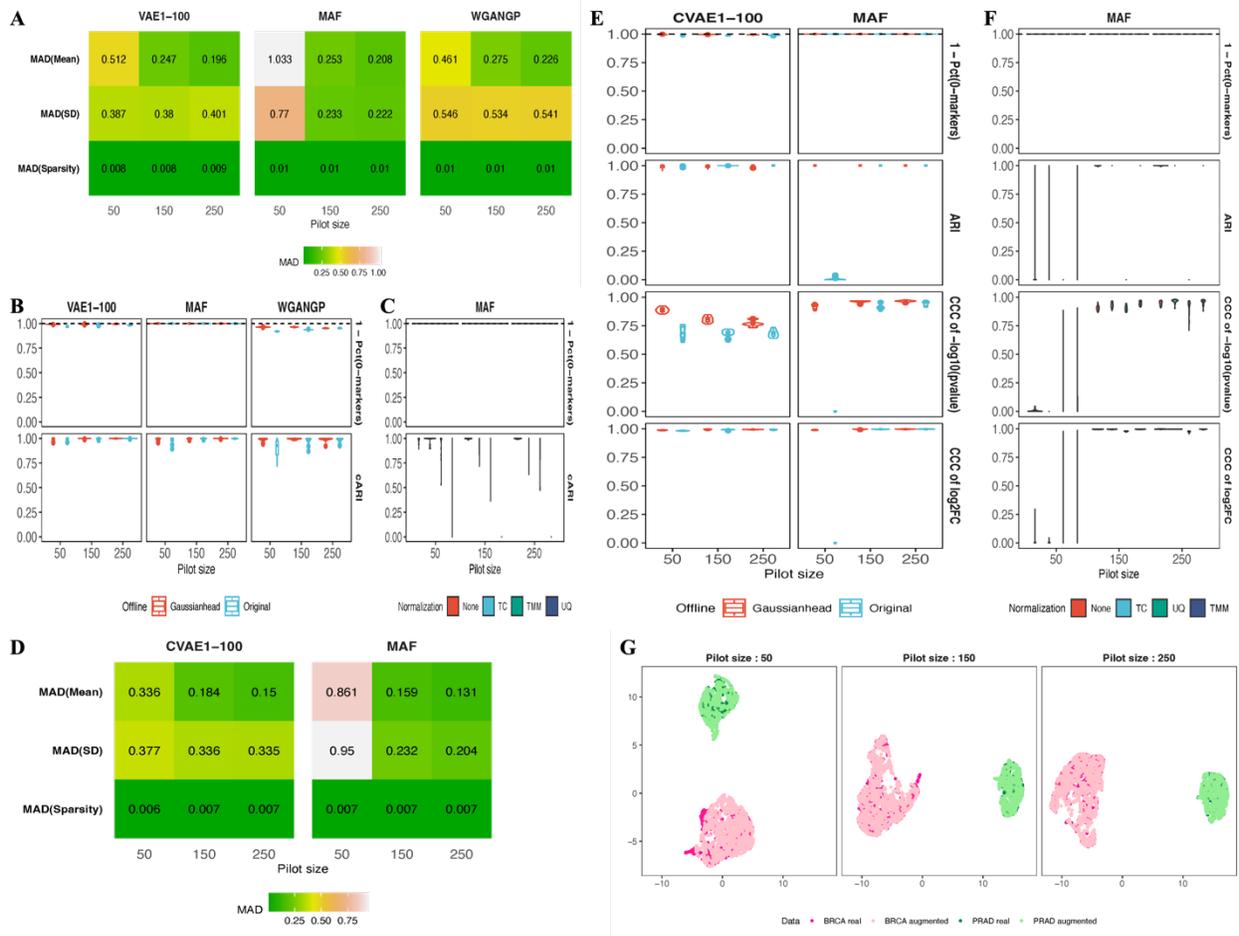

Fig. 4 | SyNG-BTS evaluation for RNA-seq in the one-group setting (Panels A-C), using pilot data from the TCGA BRCA study, and in the two-group setting (Panels D-G), using pilot data from the combination of the TCGA BRCA and PRAD studies, both with marker filtering. A: MADs in marker-specific summary statistics (mean, standard deviation, and sparsity) between the augmented data and the empirical data are calculated as the pilot data sample size increases from 50 to 250. B: Additional evaluation metrics for data congruence, including 1 – Pct(0-markers) and cARI, are calculated over varying pilot data sample sizes, with or without offline augmentation via Gaussian noise addition (indicated by colors). C: Evaluation metrics for data congruence, including 1 – Pct(0-markers) and cARI, are calculated using three different depth normalization methods for pilot data (indicated by colors) in comparison with no normalization. D: MADs in marker-specific summary statistics between the augmented data and the empirical data, are calculated as the pilot data sample size increases from 50 to 250 per sample group. E: Evaluation metrics for data congruence, including 1 – Pct(0-markers) and ARI, are calculated with or without the use of offline augmentation via Gaussian noise addition (indicated by colors). F: Evaluation metrics for data congruence are calculated using three different depth normalization methods for pilot data (indicated by colors) in comparison with no normalization. G: The

Uniform Manifold Approximation and Projection (UMAP) representation for the generated samples (by CVAE1-100) and the real samples for varying pilot data sample sizes, with the data source and the sample type indicated by colors.

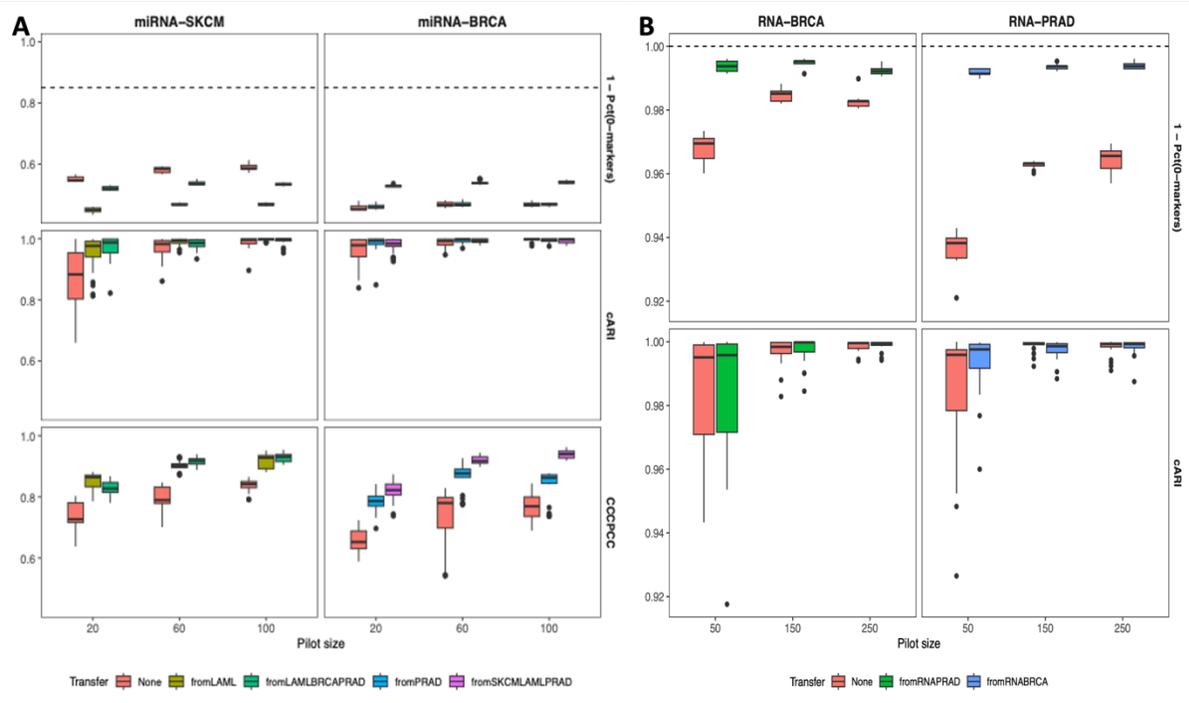

**Fig. 5 | Evaluation of transfer learning for enhancing model training in SyNG-BTS using microRNA-seq data (Panel A) and RNA-seq data (Panel B) with marker filtering. A**: Evaluation metrics on the congruence of the augmented data and the empirical data, including 1 – Pct(0-markers), cARI, and CCCPCC, are calculated when pilot data are drawn from the TCGA SKCM microRNA-seq study and models are pre-trained using the TCGA LAML study or the combination of the TCGA BRCA, LAML, and PRAD studies (left column of sub-panels), and when pilot data are drawn from the TCGA BRCA microRNA-seq study and the models are pre-trained using the TCGA PRAD or the combination of the TCGA SKCM, LAML, and PRAD studies (right column of sub-panels). **B**: Evaluation metrics for data congruence, including 1 – Pct(0-markers) and cARI, are calculated when pilot data are drawn from the TCGA BRCA RNA-seq study and the models are pre-trained using the TCGA PRAD study (left column of sub-panels), and when pilot data are drawn from the TCGA PRAD RNA-seq study and the models are pre-trained using the TCGA BRCA study (right column of sub-panels).

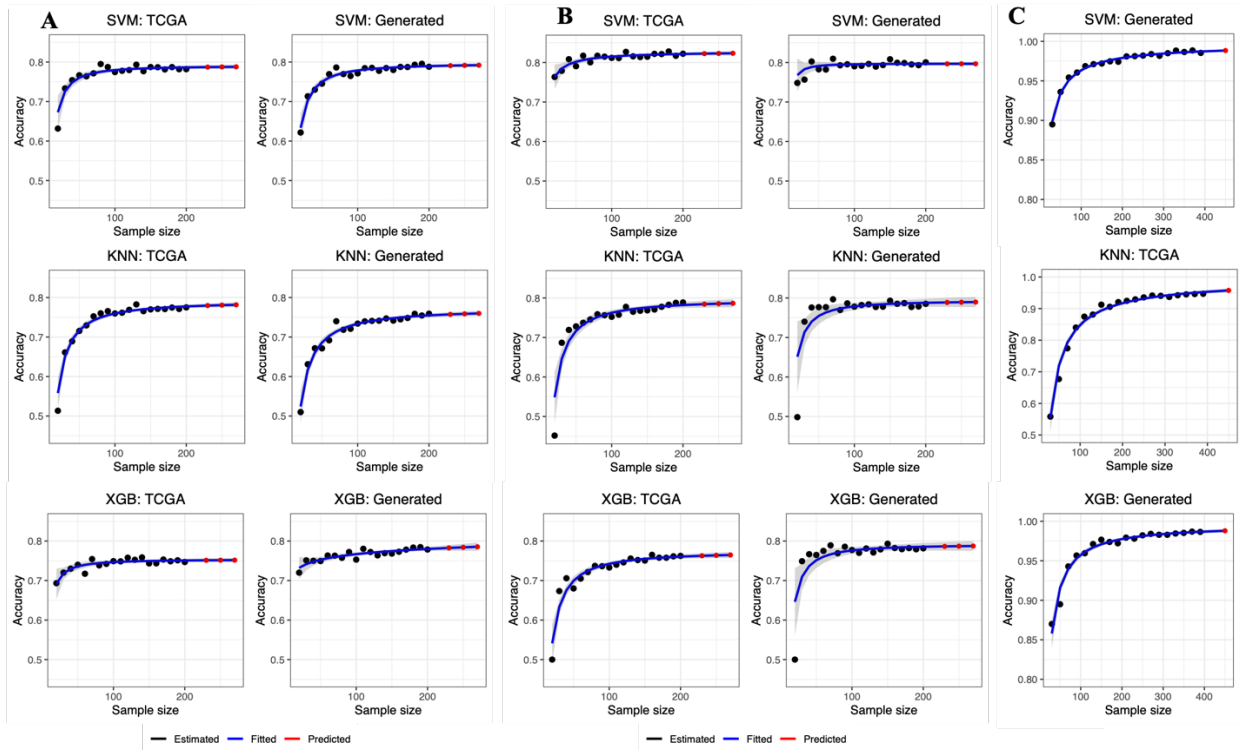

**Fig. 6 | Evaluation of SyntheSize on microRNA-seq data (Panel A) and RNA-seq data (Panel B) from the TCGA BRCA study, and application of SyntheSize to RNA-seq data from a clinical study of nivolumab (Panel C). A and B**: Classifiers are constructed to distinguish the two breast cancer subtypes, Invasive Ductal Carcinoma (IDC) and Invasive Lobular Carcinoma (ILC), in the TCGA BRCA study using empirical data (left column in each panel) or the SyNG-BTS augmented data (right column in each panel) and employing three machine learning techniques (top row in each panel: Support Vector Machine [SVM]; middle row: K-Nearest Neighbors [KNN]; bottom row: XGBoost [XGB]). **C**: Classifiers are built to predict patient response to nivolumab, Complete/Partial Response and Progressive/Stable Disease, using RNA-seq data from a published clinical study as pilot data. In panels A-C, classification accuracies are assessed for three learning techniques, including SVM, KNN, and XGB, across a range of sample sizes. Specifically, classification accuracies estimated from empirical or augmented data are plotted as black dots, while their fitted IPLFs are plotted as blue curves, projecting accuracies achieved at additional sample sizes indicated by red dots. The gray bands represent the 95% confidence regions for the fitted IPLFs.

# Extended Data for "Optimizing Sample Size for Supervised Machine Learning in Bulk Transcriptomic Sequencing: A Learning Curve Approach"

Yunhui Qi[1,2], Xinyi Wang[1,3], Li-Xuan Qin[1,*]

[1] Department of Epidemiology and Biostatistics, Memorial Sloan Kettering Cancer Center, New York, NY, United States
[2] Department of Statistics, Iowa State University, Ames, IA, United States
[3] Department of Statistics, The University of California, Davis, CA, United States
* Corresponding Author

**Supplementary Figures: S1-S13**

**Supplementary Methods**

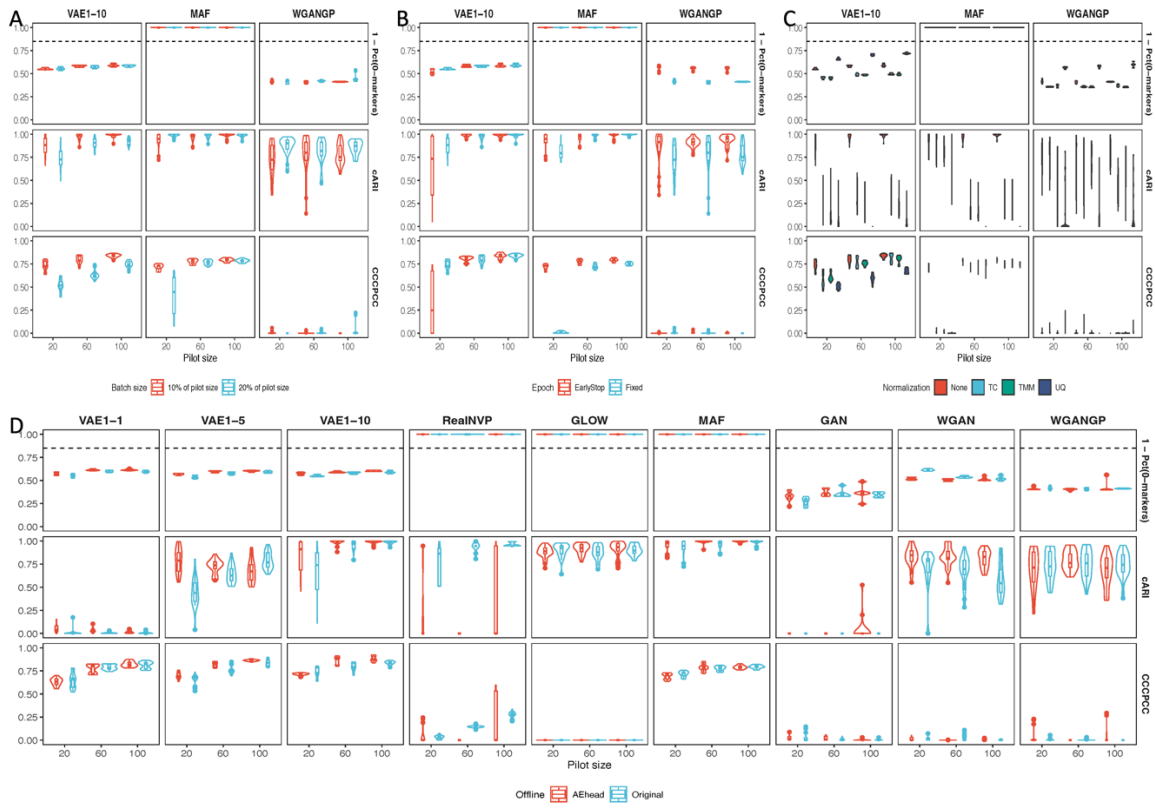

**Fig. S1 | SyNG-BTS evaluation for microRNA-seq in the one-group setting, using pilot data from the TCGA SKCM study without marker filtering.** **A**: Evaluation metrics for assessing the congruence between the SyNG-BTS augmented data and the empirical data, including (1) the percentage of markers with non-zero counts in at least one sample (indicated as 1 – Pct(0-markers)), (2) the agreement of sample clusters and data sources when clustering a combined dataset of both generated and real samples, measured by the complementary Adjusted Rand Index (cARI), and (3) the degree of correlation among member microRNAs belonging to the same polycistronic clusters, quantified by the Concordance Correlation Coefficient of Partial Correlation Coefficients (CCCPCC), are calculated for the best performing variant in each generative model family as the pilot data sample size increases from 20 to 100, using two different training batch sizes (indicated by colors). **B**: Evaluation metrics for data congruence, including 1 – Pct(0-markers), cARI, and CCCPCC, are calculated using two different epoch strategies (indicated by colors). **C**: Evaluation metrics for data congruence, including 1 – Pct(0-markers), cARI, and CCCPCC, are calculated using three different normalization methods (indicated by colors): Total Count (TC), Trimmed Mean of M-values (TMM), and Upper Quartile (UQ), in comparison with no normalization (None) for pilot data. **D**: Evaluation metrics for data congruence, including 1 – Pct(0-markers), cARI, and CCCPCC, are presented with or without the use of offline augmentation via AE head (indicated by colors).

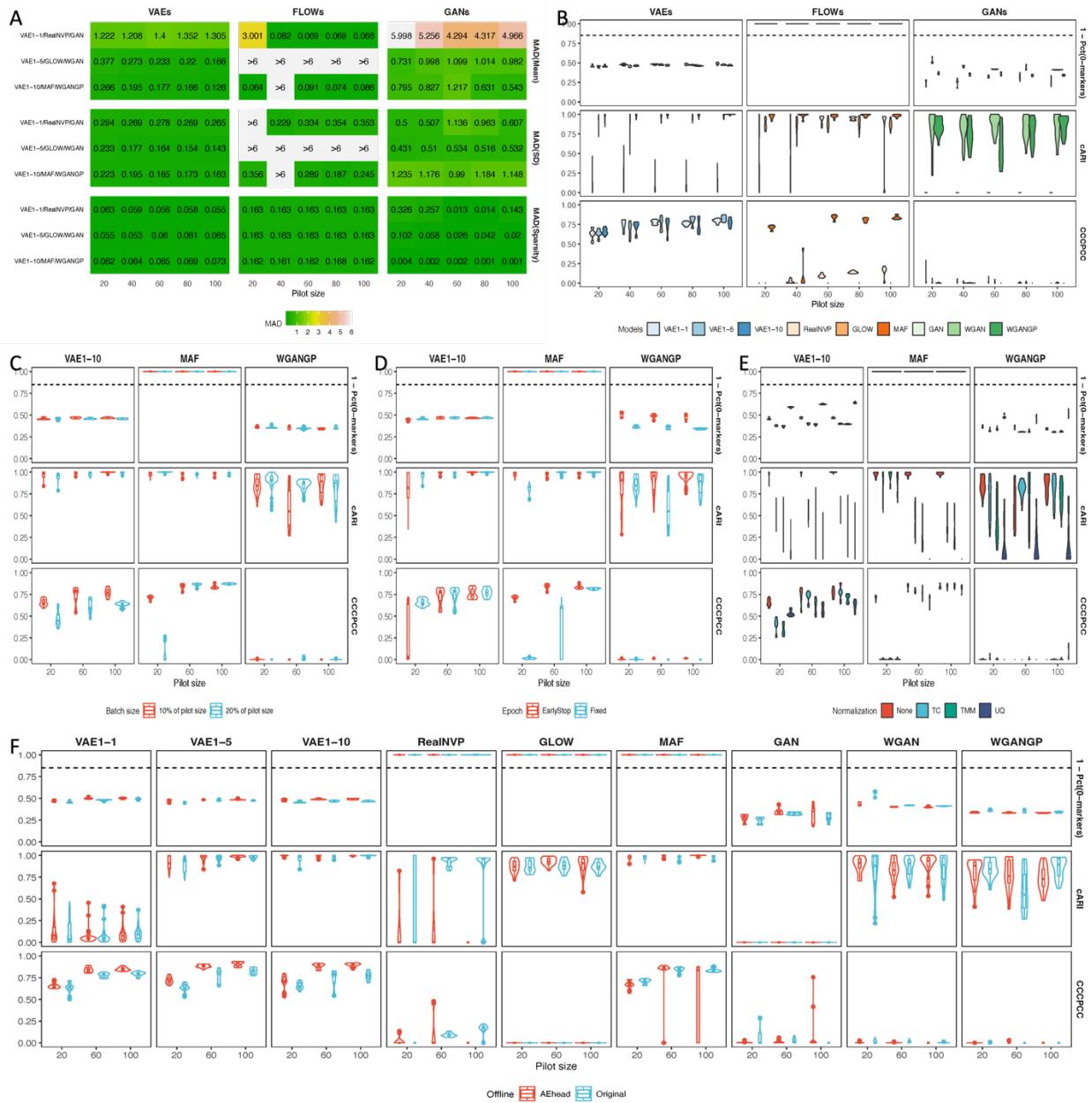

**Fig. S2 | SyNG-BTS evaluation for microRNA-seq in the one-group setting, using pilot data from the TCGA BRCA study without marker filtering.** **A**: Median Absolute Deviations (MADs) in marker-specific summary statistics (mean, standard deviation, and sparsity, defined as the percentage of zeros) between the SyNG-BTS augmented data and the empirical data are calculated as the pilot data sample size increases from 20 to 100. The MAD values are color-coded, with extremely large values represented as ">6". Smaller MADs indicate better congruency between the augmented data and the empirical data. Each sub-panel column represents one of the three generative model families, and each row within a sub-panel corresponds to a specific model variant, as indicated on the left of each sub-panel. **B**: Additional evaluation metrics assessing the congruence between the SyNG-BTS augmented data and the empirical data, encompassing 1 – Pct(0-markers), cARI, and CCCPCC, are calculated for various pilot data sample sizes. **C**: Evaluation metrics for data congruence, including 1 – Pct(0-markers), cARI, and CCCPCC, are calculated for the best performing variant in each generative model family, using two different training batch sizes (indicated by colors). **D**: Evaluation metrics for data congruence are calculated using two

different epoch strategies (indicated by colors). **E**: Evaluation metrics for data congruence are calculated using three different depth normalization methods (indicated by colors), in comparison with no normalization for pilot data. **F**: Evaluation metrics for data congruence are calculated with or without the use of offline augmentation via AE head (indicated by colors).

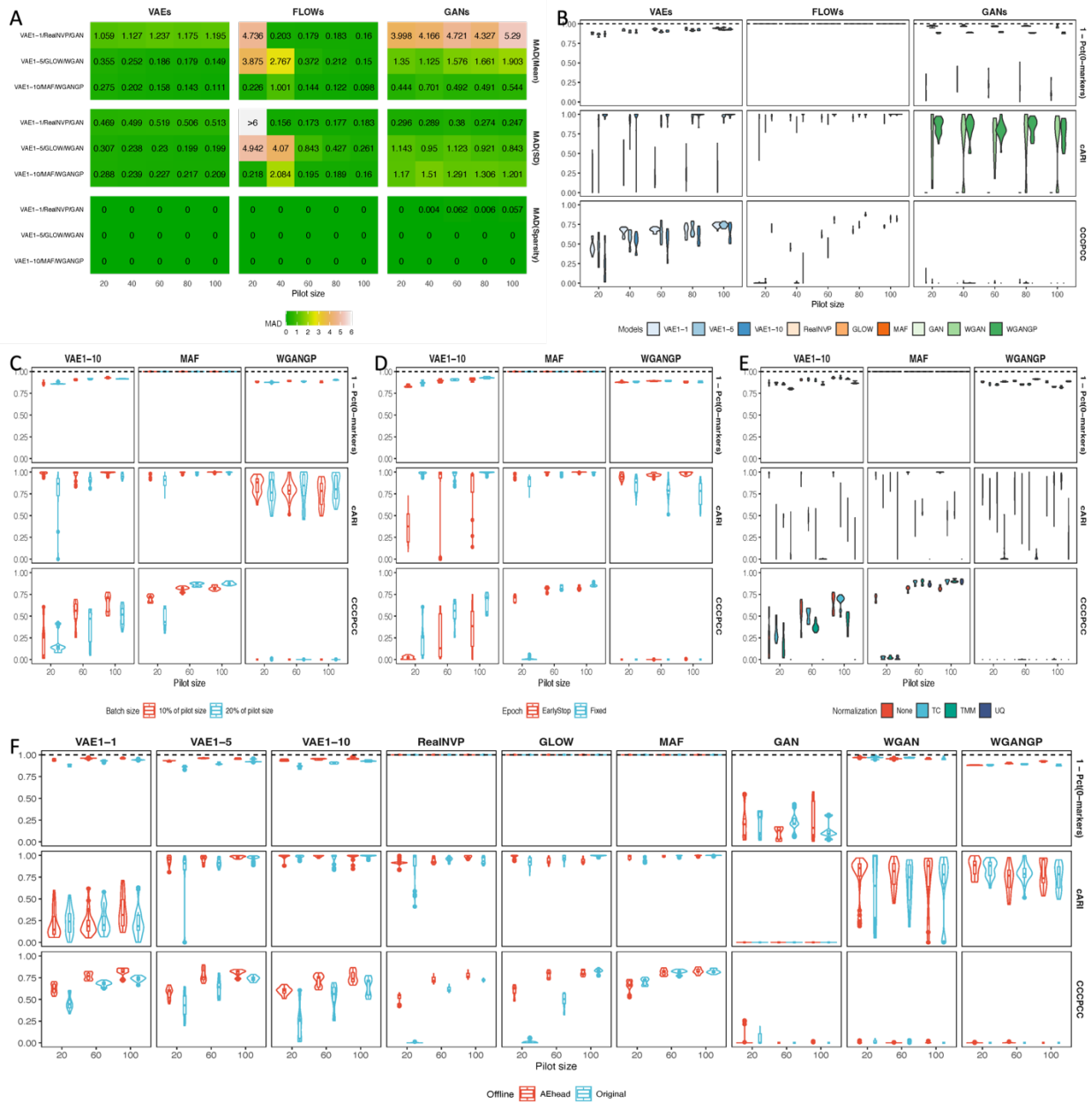

**Fig. S3 | SyNG-BTS evaluation for microRNA-seq in the one-group setting, using pilot data from the TCGA BRCA study with marker filtering. A**: MADs in marker-specific summary statistics (mean, standard deviation, and sparsity) between the SyNG-BTS augmented data and the empirical data are calculated as the pilot data sample size increases from 20 to 100. The MAD values are color-coded, with extremely large values represented as ">6". Smaller MADs indicate better congruency between the augmented data and the empirical data. Each sub-panel column represents one of the three generative model families, and each row within a sub-panel corresponds to a specific model variant, as indicated on the left of each sub-panel. **B**: Additional evaluation metrics assessing data congruence, encompassing 1 – Pct(0-markers), cARI, and CCCPCC, are calculated across various pilot data sample sizes. **C**: Evaluation metrics for data congruence, including 1 – Pct(0-markers), cARI, and CCCPCC, are calculated for the best performing variant in each generative model family, using two different training batch sizes (indicated by colors). **D**: Evaluation metrics for data congruence are calculated using two different epoch strategies (indicated by colors). **E**: Evaluation metrics for data congruence are calculated using three different depth

normalization methods (indicated by colors), in comparison with no normalization for pilot data. **F**: Evaluation metrics for data congruence are calculated with or without the use of offline augmentation via AE head (indicated by colors).

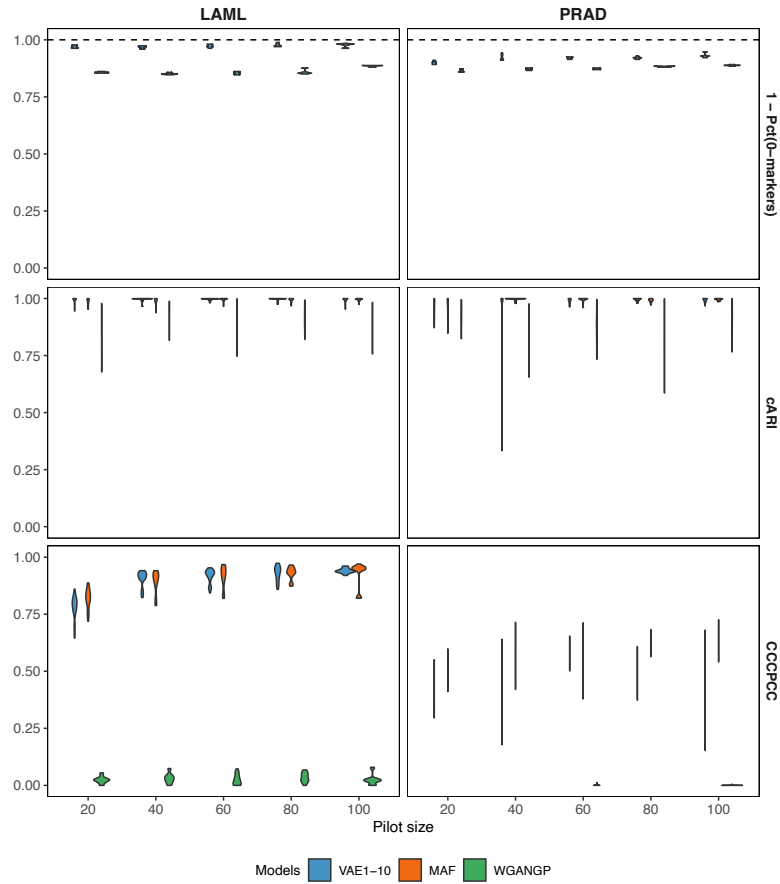

**Fig. S4 | SyNG-BTS evaluation for microRNA-seq in the one-group setting, using pilot data from the TCGA LAML (left panels) and PRAD (right panels) studies with marker filtering, employing selected generative models (represented by different colors) under a preferred training setting (that is, with the use of offline augmentation, 10% batch fraction, fixed epochs for VAE1-10, and early stopping for MAF and WGANGP).** Evaluation metrics assessing the congruence between the SyNG-BTS augmented data and the empirical data, including 1 – Pct(0-markers), cARI, and CCCPCC, are calculated as the pilot data sample size increases from 20 to 100.

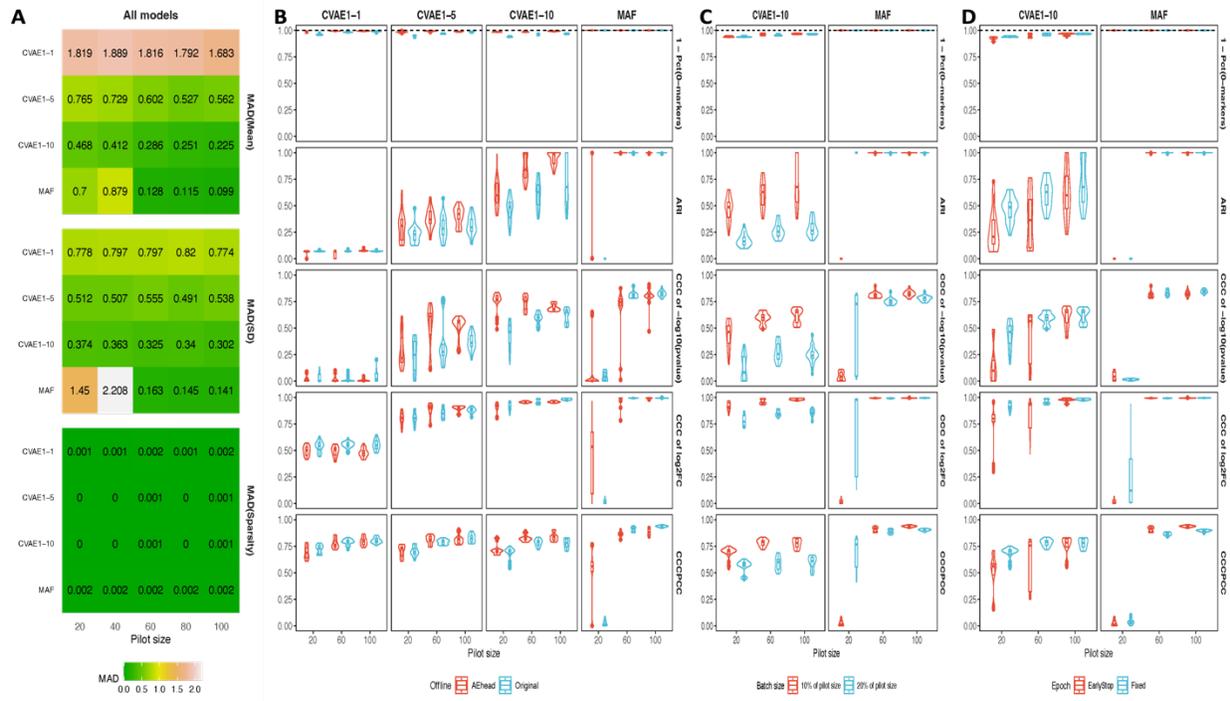

**Fig. S5 | SyNG-BTS evaluation for microRNA-seq in the two-group setting, using pilot data from the combination of the TCGA SKCM and LAML studies with marker filtering**. **A**: MADs in marker-specific summary statistics (mean, standard deviation, and sparsity) between the SyNG-BTS augmented data (using CVAE with various loss ratios and MAF) and the empirical data are calculated as the pilot data sample size per sample group increases from 20 to 100. **B**: Additional evaluation metrics assessing the congruence between the augmented data and the empirical data, encompassing (1) 1 – Pct(0-markers), (2) the agreement of sample clusters and sample types when clustering a combined dataset of both generated and real samples, measured by the Adjusted Rand Index (indicated as ARI), (3) concordant correlation coefficient of p-values from differential expression analysis on the –log10 scale (indicated as CCC of –log10 (p-value)), (4) concordant correlation coefficient of log2 fold change from differential expression analysis (indicated as CCC of log2FC), and (5) CCCPCC, are calculated with or without the use of offline augmentation (indicated by colors). **C**: Evaluation metrics for data congruence are calculated using two different training batch sizes (indicated by colors). **D**: Evaluation metrics for data congruence are calculated using two different epoch strategies (indicated by colors).

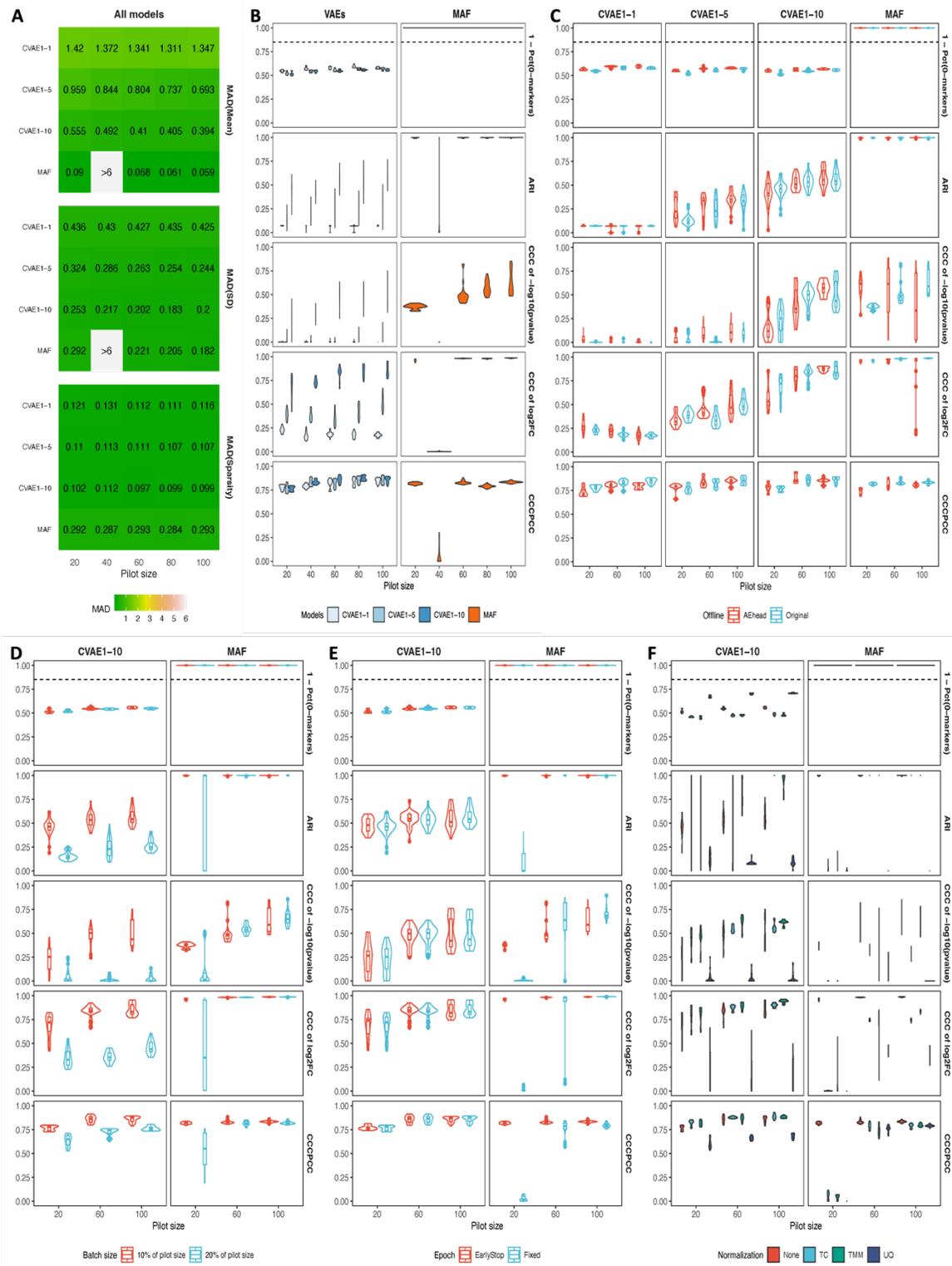

**Fig. S6 | SyNG-BTS evaluation for microRNA-seq in the two-group setting, using pilot data from the combination of the TCGA SKCM and LAML studies without marker filtering**. **A**: MADs in marker-specific summary statistics (mean, standard deviation, and sparsity) between the SyNG-BTS augmented

data (using CVAE with various loss ratios and MAF) and the empirical data are calculated as the pilot data sample size per group increases from 20 to 100. **B**: Additional evaluation metrics assessing the congruence between the augmented data and the empirical data, encompassing 1 – Pct(0-markers), ARI, CCC of –log10 (p-value), CCC of log2FC, and CCCPCC, are calculated for various pilot data sample sizes. **C**: Evaluation metrics for data congruence are calculated with or without the use of offline augmentation (indicated by colors). **D**: Evaluation metrics for data congruence are calculated using two different training batch sizes (indicated by colors). **E**: Evaluation metrics for data congruence are calculated using two different epoch strategies (indicated by colors). **F**: Evaluation metrics for data congruence are calculated using three different depth normalization methods (indicated by colors), in comparison with no normalization for pilot data.

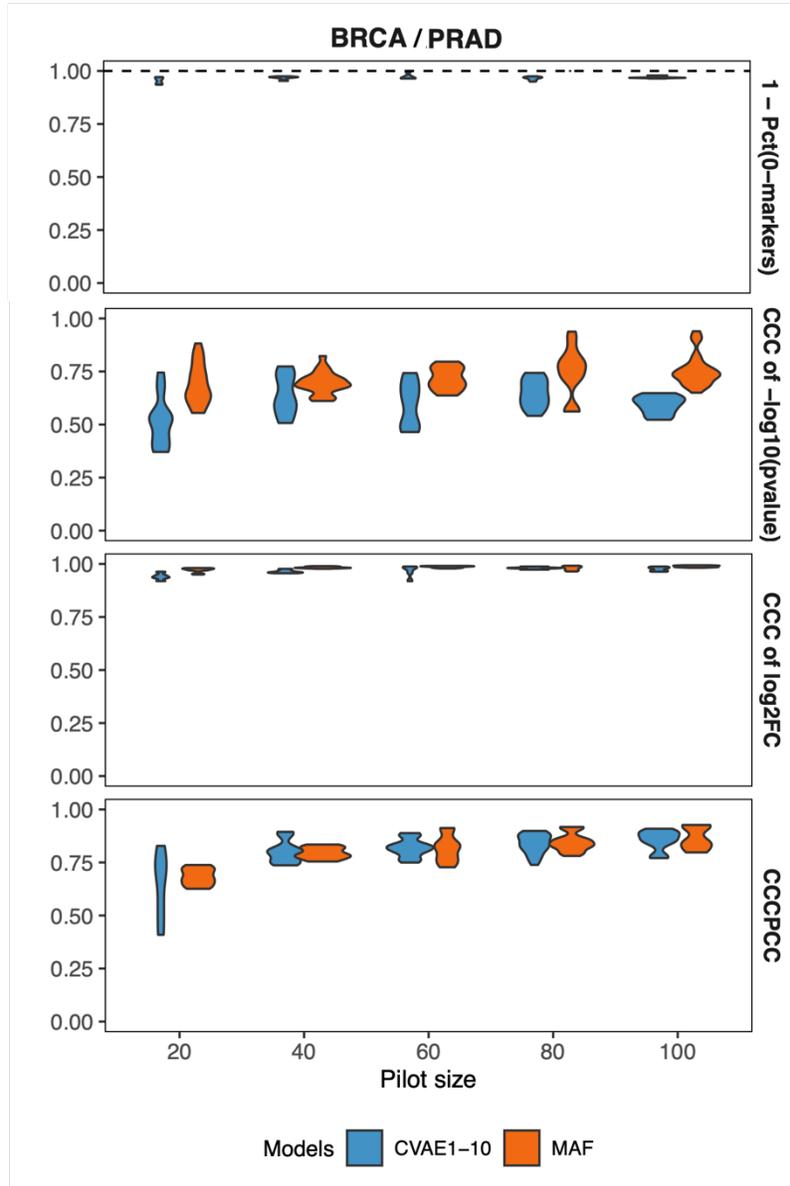

**Fig. S7 | SyNG-BTS evaluation for microRNA-seq in the two-group setting, using pilot data from the combined TCGA BRCA and PRAD studies with marker filtering, employing selected generative models under a preferred training setting (that is, with the use of offline augmentation, 10% batch fraction, fixed epochs for CVAE1-10, and early stopping for MAF).** Evaluation metrics for data congruence, including 1 – Pct(0-markers), ARI, CCC of –log10 (p-value), CCC of log2FC, and CCCPCC, are calculated as the pilot data sample size per sample group increases from 20 to 100.

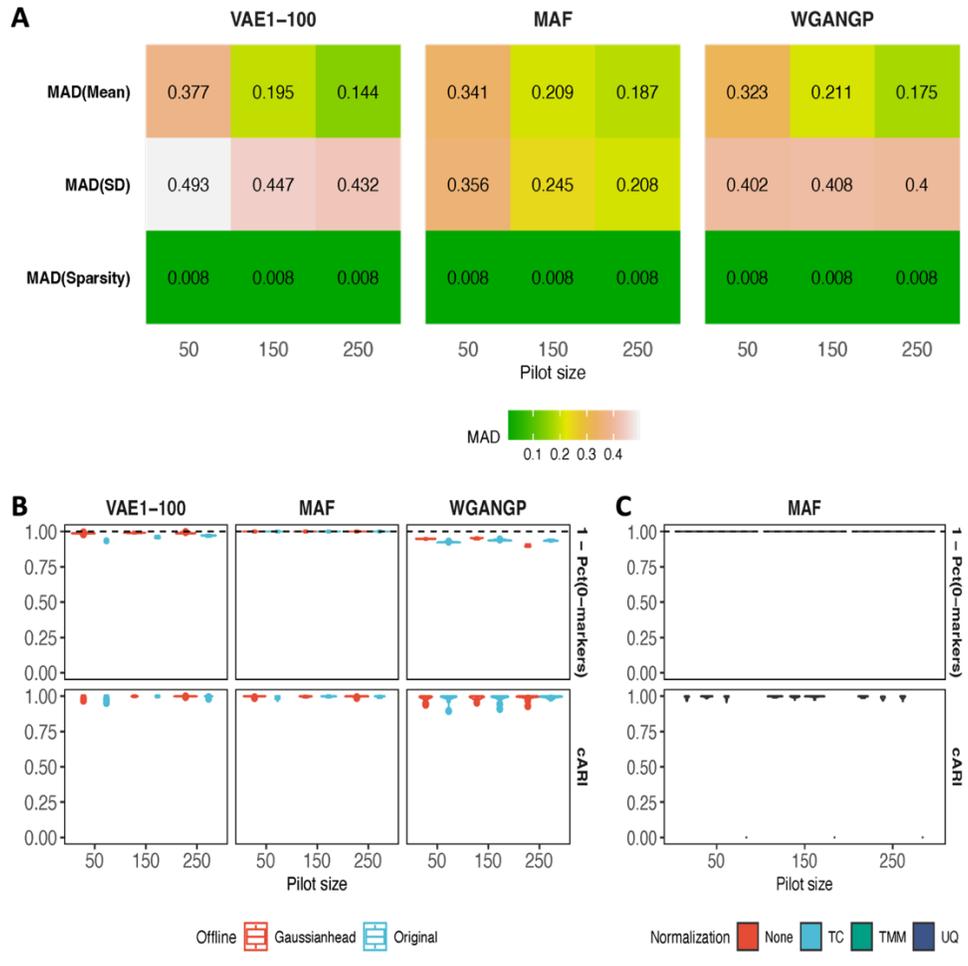

**Fig. S8 | SyNG-BTS evaluation for RNA-seq in the one-group setting, using pilot data from the TCGA PRAD study with marker filtering.** **A**: MADs in marker-specific summary statistics (sub-panel rows) between the SyNG-BTS augmented data and the empirical data are calculated, as the pilot data sample size increases from 50 to 250. **B**: Additional evaluation metrics assessing data congruence, encompassing 1 − Pct(0-markers) and cARI, are calculated with or without the use of offline augmentation via Gaussian noise addition (indicated by colors). **C**: Evaluation metrics for data congruence are calculated using three different depth normalization methods (indicated by colors), in comparison with no normalization for pilot data.

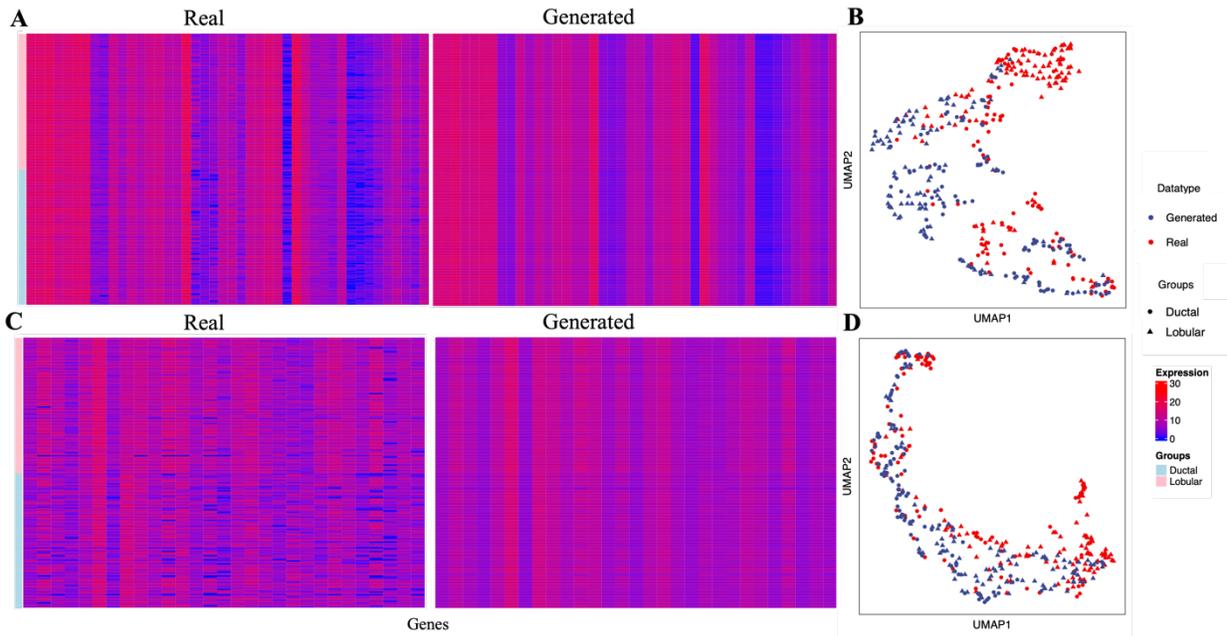

**Fig. S9 | Comparison of the generated samples with the real samples from the TCGA BRCA study: miRNA-seq (panels A and B) and RNA-seq (panels C and D). A and C**: Heatmap of the data for the real samples (left sub-panel) and the generated samples (right sub-panel), on the log2 scale. **B and D**: UMAP of the data for the real samples and the generated sample, with the data source indicated by colors and the sample type indicated by point shapes.

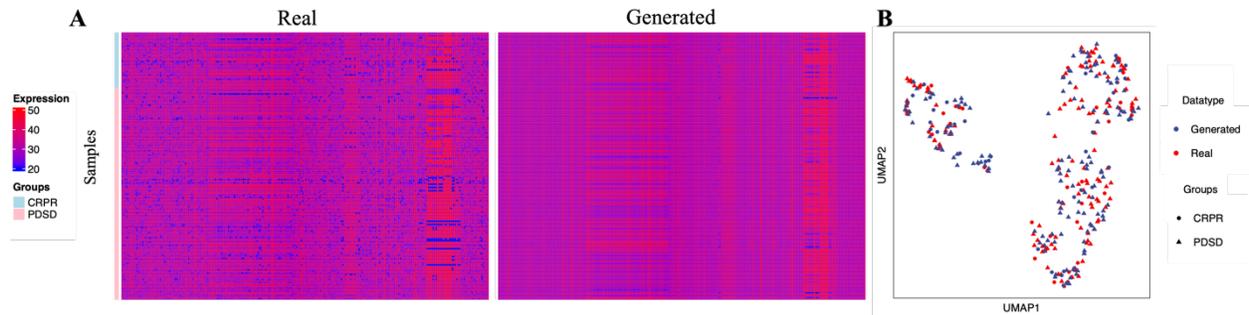

**Fig. S10 | Evaluation of SyNG-BTS generated samples based on pilot RNA-seq data taken from a clinical study of nivolumab in advanced clear cell renal cell carcinoma (Braun et al., 2020)[1].** **A**: Heatmap of the data for real samples (left sub-panel) and generated samples (right sub-panel), on the log2 scale. **B**: UMAP of the data for the real samples and the generated sample, with the data source indicated by colors and the patient response status (Complete/Partial response [CRPR]) *versus* Progressive/Stable Disease [PDSD]) indicated by point shapes.

## Supplementary Methods: Review of Deep Generative Models

We review the principle of three categories of DGMs and describe the neural net structures used in our study (Fig. S11-13).

**Autoencoder (AE).** AE was first introduced by [2] as a neural network architecture designed for learning a compact, low-dimensional representation of data. This is achieved by minimizing the reconstruction loss, typically measured as the mean squared error between the input samples and their corresponding reconstructed data. Comprising an encoder and a decoder, both equipped with non-linear activation functions, AEs perform non-linear transformations. The encoder converts the original data into a lower-dimensional representation, while the decoder reconstructs the data from this representation. By employing non-linear activation functions in both components, AEs effectively project data onto a well-fitting non-linear manifold. However, a drawback of traditional AEs is that the learned low-dimensional representation is a fixed function of the data, making it challenging to interpolate. In response to this limitation, the variational AE was introduced to offer a more flexible and probabilistic approach to encoding data.

**Variational Autoencoder (VAE).** Like AE, VAE [3] also has an encoder and a decoder. But instead of learning a fixed function as the low dimensional representation, it learns a random distribution. The fitting of VAE is to optimize the log-likelihood of the data by maximizing the evidence lower bound (ELBO).

Suppose we have data $\mathbf{x} \in \mathbb{R}^{D_x}$, we would like to learn the distribution of a low dimensional random vector $\mathbf{z} \in \mathbb{R}^{D_z}$, which has density function $q_\phi(\mathbf{z}|\mathbf{x})$ indexed with unknown parameters $\phi$. The encoder aims to transform the observed $\mathbf{x}$ into $\mathbf{z}$ through $q_\phi(\mathbf{z}|\mathbf{x})$. Similarly, we define the decoder transformation as $p_\theta(\mathbf{x}|\mathbf{z})$ with unknown parameters $\theta$. The training is to maximize the log-likelihood of the data $\log p(\mathbf{x})$. Denote the joint density function of $\mathbf{x}$ and $\mathbf{z}$ as $p(\mathbf{x}, \mathbf{z})$, the conditional distribution of $\mathbf{x}$ given $\mathbf{z}$ as $p(\mathbf{x}|\mathbf{z})$, the density of $\mathbf{z}$ as $p(\mathbf{z})$, the conditional distribution of $\mathbf{z}$ given $\mathbf{x}$ as $q(\mathbf{z}|\mathbf{x})$. We have the ELBO as the lower bound for log likelihood of data.

$$\begin{aligned}
\log p(\mathbf{x}) &= \log \int p(\mathbf{x}, \mathbf{z}) \\
&= \log \int p(\mathbf{x}, \mathbf{z}) \frac{q(\mathbf{z}|\mathbf{x})}{q(\mathbf{z}|\mathbf{x})} \\
&= \log \int \frac{p(\mathbf{x}, \mathbf{z})}{q(\mathbf{z}|\mathbf{x})} q(\mathbf{z}|\mathbf{x}) \\
&= \log \mathbb{E}_{\mathbf{z} \sim q(\mathbf{z}|\mathbf{x})} \frac{p(\mathbf{x}, \mathbf{z})}{q(\mathbf{z}|\mathbf{x})} \\
&\geq \mathbb{E}_{\mathbf{z} \sim q(\mathbf{z}|\mathbf{x})} \log \frac{p(\mathbf{x}, \mathbf{z})}{q(\mathbf{z}|\mathbf{x})} \\
&= \mathbb{E}_{\mathbf{z} \sim q(\mathbf{z}|\mathbf{x})} \log \frac{p(\mathbf{x}|\mathbf{z}) p(\mathbf{z})}{q(\mathbf{z}|\mathbf{x})} \\
&= \mathbb{E}_{\mathbf{z} \sim q(\mathbf{z}|\mathbf{x})} \log p(\mathbf{x}|\mathbf{z}) - \int q(\mathbf{z}|\mathbf{x}) \log \frac{q(\mathbf{z}|\mathbf{x})}{p(\mathbf{z})} d\mathbf{z} \\
&= \mathbb{E}_{\mathbf{z} \sim q(\mathbf{z}|\mathbf{x})} \log p(\mathbf{x}|\mathbf{z}) - \mathrm{KL}[q(\mathbf{z}|\mathbf{x}) || p(\mathbf{z})] \\
&= \mathrm{ELBO}.
\end{aligned}$$

The optimization of ELBO is through stochastic gradient descent (SGD) algorithm. To do this, we need an explicit form of the ELBO. The second term of ELBO - KL-divergence can be computed by assuming

$$q_\phi(\mathbf{z}|\mathbf{x}) = N\left(\boldsymbol{\mu}_\phi(\mathbf{x}), \boldsymbol{\Sigma}_\phi(\mathbf{x})\right),$$
$$p(\mathbf{z}) = N(\mathbf{0}, \mathbf{I}_{D_z});$$

Then,

$$\mathrm{KL}[q(\mathbf{z}|\mathbf{x})||p(\mathbf{z})] = \frac{1}{2}\left[\mathrm{tr}\boldsymbol{\Sigma}_\phi(\mathbf{x}) - D_z - \log|\boldsymbol{\Sigma}_\phi(\mathbf{x})| + \boldsymbol{\mu}_\phi(\mathbf{x})^T\boldsymbol{\mu}_\phi(\mathbf{x})\right].$$

For the first part of ELBO - $\mathbb{E}_{z\sim q(z|x)}\log p(\mathbf{x}|\mathbf{z})$, Monte Carlo approximation provides a better option. Approximate $\mathbb{E}_{z\sim q(z|x)}\log p(\mathbf{x}|\mathbf{z})$ by

$$\frac{1}{M}\sum_{i=1}^{M}\log p_\theta(\mathbf{x}|\mathbf{z}^i),$$

where $\mathbf{z}^i, i = 1, \ldots, M$ are samples from $q_\theta(\mathbf{z}|\mathbf{x})$.

With assumption $p_\theta(\mathbf{x}|\mathbf{z}) = N(\boldsymbol{\mu}_\theta(\mathbf{z}), \sigma^2 \mathbf{I}_{D_x})$, we have

$$\mathbb{E}_{z\sim q(z|x)}\log p(\mathbf{x}|\mathbf{z}) \approx \frac{1}{M}\sum_{i=1}^{M}\log p_\theta(\mathbf{x}|\mathbf{z}^i) = \frac{1}{M}\sum_{i=1}^{M}\left[C - \frac{1}{2\sigma^2}\left(\mathbf{x} - \boldsymbol{\mu}_\theta(\mathbf{z}^i)\right)^T\left(\mathbf{x} - \boldsymbol{\mu}_\theta(\mathbf{z}^i)\right)\right]$$

where $C$ is a constant with respect to parameters $\theta$. Notice that $\theta$ is the parameter we want to optimize, therefore, only the part $-\frac{1}{\sigma^2}\frac{1}{M}\sum_{i=1}^{M}\left(\mathbf{x} - \boldsymbol{\mu}_\theta(\mathbf{z}^i)\right)^T\left(\mathbf{x} - \boldsymbol{\mu}_\theta(\mathbf{z}^i)\right)$ matters. Equivalently, we can minimize the weighted mean squared distance between the data $\mathbf{x}$ and the reconstructed data $\boldsymbol{\mu}_\theta(\mathbf{z}^i)$ with weights $1/\sigma^2$.

However, it is difficult to sample $\mathbf{z}^i$ from an intermediate layer of the whole VAE, a reparameterization trick is introduced to move the sampling at inner layer to the input layer. Instead of sampling $\mathbf{z}^i$ from $N\left(\boldsymbol{\mu}_\phi(\mathbf{x}), \boldsymbol{\Sigma}_\phi(\mathbf{x})\right)$, sample $\boldsymbol{\varepsilon}^i$ from $N(\mathbf{0}, \mathbf{I}_{D_z})$, and we can get $\mathbf{z}^i$ by

$$\mathbf{z}^i = \boldsymbol{\mu}_\phi(\mathbf{x}) + \boldsymbol{\Sigma}_\phi(\mathbf{x})^{1/2}\boldsymbol{\varepsilon}^i.$$

We can get the loss function of VAE $\mathbf{z}^i = \boldsymbol{\mu}_\phi(\mathbf{x}) + \boldsymbol{\Sigma}_\phi(\mathbf{x})^{1/2}\boldsymbol{\varepsilon}^i$

$$\begin{aligned}\log p(\mathbf{x}) &\approx \mathrm{ELBO}(\theta, \phi)\\ &\approx \text{reconstruction-loss} + \text{KL-divergence}\\ &= -\frac{1}{\sigma^2}\left[\frac{1}{M}\sum_{i=1}^{M}\left(\mathbf{x} - \boldsymbol{\mu}_\theta(\mathbf{z}^i)\right)^T\left(\mathbf{x} - \boldsymbol{\mu}_\theta(\mathbf{z}^i)\right)\right]\\ &\quad + \left[\mathrm{tr}\boldsymbol{\Sigma}_\phi(\mathbf{x}) - \log|\boldsymbol{\Sigma}_\phi(\mathbf{x})| + \boldsymbol{\mu}_\phi(\mathbf{x})^T\boldsymbol{\mu}_\phi(\mathbf{x})\right].\end{aligned}$$

With fixed $\mathbf{x}, \mathbf{z}$, the loss function is continuous in $\theta$ and $\phi$, therefore we can use SGD to find the solution.

Since VAE learns a random distribution instead of a fixed function of the low dimensional representation, we can sample from the distribution of $\mathbf{z}$ to generate more data. This is how VAE performs as a generative model. However, VAE is an unsupervised learning model. Once the dataset includes samples from different groups, VAE is not directly applicable to the whole dataset since it is not reasonable to assume the samples from different groups have the same underlying distribution. One solution is that we can apply VAE to samples from each group. But we might lose power since each group have smaller sample size. Conditional VAE was introduced to handle datasets with multiple groups, or in other words, labels.

To incorporate the group information of samples, we use conditional VAE (CVAE). The structure of CVAE is essentially the same as VAE. Both include the encoder and decoder. However, to take the groups or labels y into consideration, the encoder in CVAE tries to learn the distribution of the low-dimensional representation $\mathbf{z}$ conditional on the data points $\mathbf{x}$ and the labels y, denoted as $p_\theta(\mathbf{z}|\mathbf{x}, y)$. Similar, the decoder of CVAE also takes the labels into consideration by learning the distribution of $\mathbf{x}$ conditional on $\mathbf{z}$ and y, denoted as $p_\phi(\mathbf{x}|\mathbf{z}, y)$. In this way, we can guide the model to specifically generate samples with the labels we want.

**Generative Adversarial Network (GAN).** Different from VAEs, the generative adversarial network (GAN) [4] provides a smart solution to generate the data, turning an unsupervised learning problem to a supervised one. It has two sub-models, the discriminator and the generator. The generator generates fake samples from noises while the discriminator learns to distinguish the empirical samples from the fake samples produced by the generator. A successfully trained GAN should reach a point where its discriminator is fooled by the fake samples approximately half of the time, indicating that the fake samples closely resemble the real ones and the discriminator cannot reliably distinguish between them.

Following the notations from [4], suppose the input noise of generator G with unknown parameters $\theta_g$ is $\mathbf{z}$ with prior distribution $p_z(\mathbf{z})$, thus the fake samples $\tilde{\mathbf{x}}$ are $G(\mathbf{z}, \theta_g)$. The discriminator D with unknown parameters $\theta_d$ takes empirical samples $\mathbf{x}$ and fake samples $\tilde{\mathbf{x}}$ as input and gives the probability of the input to be empirical samples. The generator and discriminator play the following minimax game:

$$\min_G \max_D \mathbb{E}_{\mathbf{x} \sim p_{data}} \log D(\mathbf{x}) + \mathbb{E}_{\mathbf{z} \sim p_z} \log[1 - D(G(\mathbf{z}))].$$

Although GAN has achieved massive success, it faces challenges in the training process. First, it is not easy to stop the training by only visualizing the loss of generator and discriminator since their losses are not meaningful. This can be a very big problem when GAN is used for a continuous dataset instead of images since it is difficult to visually determine whether the fake samples are like the real ones in the continuous dataset. In addition, the training of GAN is not stable. Because of the adversarial training, it is difficult for the generator and the discriminator to converge simultaneously. And it is easy for GAN to only learn a specific mode of the underlying data distribution and generate fake samples that are not representative of the whole distribution of the real data. This is called mode collapse. [5] pointed out that these problems are from the loss function of GAN. Instead, they proposed to use Wasserstein distance.

**WGAN.** Essentially, the training of GAN is to minimize the distance between the distribution of empirical samples $p_r$ and the distribution of fake samples $p_g$. The original GAN uses Jensen–Shannon divergence to evaluate the distribution between $p_r$ and $p_g$. And this is not suitable when there are disjoint parts of $p_r$ and $p_g$. [5] theoretically analyzed several distance measures and proposed to use Wasserstein distance instead, resulting WGAN.

WGAN aims to optimize the following loss function:

$$\max_{w \in W} \mathbb{E}_{\mathbf{x} \sim p_r} f_w(\mathbf{x}) - \mathbb{E}_{\mathbf{z} \sim p_g} f_w[G(\mathbf{z})].$$

As opposed to the discriminator in GAN which is to learn to distinguish the real and fake samples, the discriminator in WGAN is to learn a K-Lipschitz continuous function $f_w$ to minimize the Wasserstein distance between $p_r$ and $p_g$. As the training progresses, the distribution learned by the generator is closer and closer to the real distribution, and the Wasserstein distance decreases. To maintain the lipschitz continuity of $f_w$ during training, the authors of WGAN proposed to use weight clipping, which means after every gradient update on the discriminator, clamp the weights w to a small, fixed range to enforce the Lipschitz continuity.

Thanks to Wasserstein distance, WGAN has a meaningful evaluation metric. By monitoring the change of Wasserstein distance, it is easy to decide when to stop the training. And the empirical study in [5] shows increased stability of training compared to GAN and no evidence of mode collapse.

WGAN is not perfect. The weight clipping to enforce the Lipschitz continuity can be problematic. When the clipping window is too large, the training can be very slow after weight clipping. When the clipping window is too small, WGAN suffers vanishing gradients. To solve this, WGAN with Gradient Penalty is proposed.

**WGANGP.** [6] proposed to add a penalty term to the loss function of WGAN to replace the weight clipping. The following is the loss function of WGAN with Gradient Penalty.

$$L = \mathbb{E}_{\tilde{\mathbf{x}} \sim p_g} D(\tilde{\mathbf{x}}) - \mathbb{E}_{\mathbf{x} \sim p_r} D(\mathbf{x}) + \lambda \mathbb{E}_{\hat{\mathbf{x}} \sim p_{\hat{x}}} \left[ \left( ||\nabla_{\hat{\mathbf{x}}} D(\hat{\mathbf{x}})||_2 - 1 \right)^2 \right],$$

where $\hat{\mathbf{x}}$ is $\epsilon \mathbf{x} + (1 - \epsilon)\tilde{\mathbf{x}}$ with $\epsilon$ as the random Gaussian noise. From the loss function of WGANGP, the only difference between WGAN and WGANGP is the penalty term which constrains the $l_2$ norm of the discriminator's gradient to be around 1. This comes from the fact that along the best optimization path, the $l_2$ norm of gradients of discriminators are always 1. Compared to WGAN, WGANGP has a more stable training process, generates high quality data, and avoids the problems brought by weight clipping.

**FLOWs**. A flow-based generative model is constructed by a sequence of invertible transformations. The flow model explicitly learns the data distribution by minimizing negative log-likelihood.

Denote probability density function of real data as $p(\mathbf{x})$, and $\mathbf{x} \in \mathcal{D}$, $\mathcal{D}$ is the domain of the real data. Given a random variable z and its probability density function $z \sim \pi(\mathbf{z})$, $\pi(\mathbf{z})$ is known and usually taken as Gaussian distribution in experiments. Construct a new random variable using a 1-1 mapping function $\mathbf{x} = f(\mathbf{z})$. The function f is invertible, so $\mathbf{z} = f^{-1}(\mathbf{x})$. The transformation of probability density function can be written as:

$$\mathbf{z} \sim \pi(\mathbf{z}), \mathbf{x} = f(\mathbf{z}), \mathbf{z} = f^{-1}(\mathbf{x})$$

$$p(\mathbf{x}) = \pi(\mathbf{z}) \frac{d\mathbf{z}}{d\mathbf{x}} = \pi(f^{-1}(\mathbf{x})) \left| \det \frac{df^{-1}}{d\mathbf{x}} \right|,$$

where det is the Jacobian determinant.

Three flow models are considered: GLOW, RealNVP, and MAF, which are all extensions of Normalizing Flows. RealNVP employs a coupling layer allowing for easy computation of the inverse and the Jacobian determinant. GLOW builds upon RealNVP by introducing 1x1 invertible convolutions, which add more flexibility and capacity. MAF considers autoregressive models for the transformation.

**Normalizing Flows.** A normalizing flow [7] transforms a simple distribution into a complex one by applying a sequence of invertible functions. Flowing through a chain of transformations, repeatedly substitute the variable for the new one according to the change of variable theorem and finally achieve the final target variable's probability distribution. Expand the output **x** step by step until tracing back to the initial distribution $\mathbf{z}_0$

$$\begin{aligned}
\mathbf{x} = \mathbf{z}_K &= f_K \circ f_{K-1} \circ \cdots \circ f_1(\mathbf{z}_0) \\
\log p(\mathbf{x}) = \log \pi_K(\mathbf{z}_K) &= \log \pi_{K-1}(\mathbf{z}_{K-1}) - \log \left| \det \frac{df_K}{d\mathbf{z}_{K-1}} \right| \\
&= \log \pi_{K-2}(\mathbf{z}_{K-2}) - \log \left| \det \frac{df_{K-1}}{d\mathbf{z}_{K-2}} \right| - \log \left| \det \frac{df_K}{d\mathbf{z}_{K-1}} \right| \\
&= \cdots \\
&= \log \pi_0(\mathbf{z}_0) - \sum_{i=1}^{K} \log \left| \det \frac{df_i}{d\mathbf{z}_{i-1}} \right|,
\end{aligned}$$

The path $\mathbf{z}_i = f_i(\mathbf{z}_{i-1})$ is the flow, and the whole chain is called a normalizing flow. To easily compute the equation, a transformation $f_i$ should satisfy two properties: (a) It's easily invertible and (b) Its Jacobian determinant is easy to compute.

Then the training criterion of the flow-based generative model is simply the negative log-likelihood (NLL) over the training dataset $\mathcal{D}$

$$\mathcal{L}(\mathcal{D}) = -\frac{1}{|\mathcal{D}|} \Sigma_{\mathbf{x} \in \mathcal{D}} \log p(\mathbf{x}).$$

**RealNVP.** The RealNVP [8] model creates a normalizing flow by stacking a sequence of invertible bijective transformation functions. In every bijective function $f: z \to x$, is defined as an affine coupling layer. In this layer, the input dimensions are split into two parts. Different transformations are applied to each part, and finally these two parts are concatenated:

$$\mathbf{z}_a, \mathbf{z}_b = \text{split}(\mathbf{z}),$$
$$(\log \mathbf{s}, \mathbf{t}) = \text{NN}(\mathbf{z}_b),$$
$$\mathbf{s} = \exp(\log \mathbf{s}),$$
$$\mathbf{x}_a = \mathbf{s} \odot \mathbf{z}_a + \mathbf{t},$$
$$\mathbf{x}_b = \mathbf{z}_b,$$
$$\mathbf{x} = \text{concat}(\mathbf{x}_a, \mathbf{x}_b).$$

where s and t are set as multilayer perceptron (MLP) with two hidden layers. It's easy to check that the transformation satisfies two basic properties for a flow transformation: (a) It's easily invertible and (b) Its Jacobian determinant is easy to compute.

**GLOW.** The GLOW [9] model extends the previous flow-based generative model, RealNVP, and simplifies the structure by replacing the reverse permutation step with invertible 1×1 convolution. There are three substeps in one step of flow in Glow.

<u>Activation normalization (actnorm).</u> It is an affine transformation using a scale and bias parameter, similar to batch normalization, but works for mini-batch with size 1. The parameters can be trained but initialized so that the first minibatch have mean 0 and standard deviation 1 after actnorm.

$$\forall i, j: \mathbf{x}_{i,j} = \mathbf{s} \odot \mathbf{z}_{i,j} + \mathbf{b}.$$

where **s** and **b** are scale and bias parameters respectively.

Invertible 1×1 convolution. For different layers of the RealNVP model, the ordering of channels is altered so that all the data dimensions have the possibility to be modified. For GLOW, a 1×1 convolution with equal number of input and output channels is a generalization of any permutation of the channel ordering

$$\forall i, j: \mathbf{x}_{i,j} = \omega_{i,j} \mathbf{z}_{i,j},$$

where $\omega_{i,j}$ is the weight.

Affine coupling layer. The design is the same as RealNVP.

**Models with Autoregressive Flows (MAF).** MAF [10] model adds the autoregressive constraint to the normalizing flow. For data $\mathbf{x} = [x_1, x_2, \cdots, x_D]$, each output only depends on the data observed in the past, but not on the future ones, i.e. the probability of observing $x_i$ is conditioned on $x_1, \cdots, x_{i-1}$ and the probability of observing the full sequence equals the product of these conditional probabilities.

$$p(\mathbf{x}) = \prod_{i=1}^{D} p(x_i \mid x_1, \ldots, x_{i-1}) = \prod_{i=1}^{D} p(x_i \mid x_{1:i-1}).$$

An autoregressive flow is a type of normalizing flow whose transformation is based on an autoregressive model. In this model, each dimension of a vector variable is conditioned on the preceding dimensions. A classic autoregressive flow model we used is Masked Autoregressive Flow (MAF). MAF is a normalizing flow with transformation layer built as an autoregressive neural network. Given two random variables, $\mathbf{z} \sim \pi(\mathbf{z})$ and $\mathbf{x} \sim p(\mathbf{x})$ and the probability density function $\pi(\mathbf{z})$ is known, MAF aims to learn $p(\mathbf{x})$. Each $x_i$ is generated conditioned on the past dimensions $x_{1:i-1}$. Like RealNVP and GLOW, MAF estimates the conditional probability by an affine transformation of $\mathbf{z}$, where the scale and shift terms are functions of the observed part of $\mathbf{x}$.

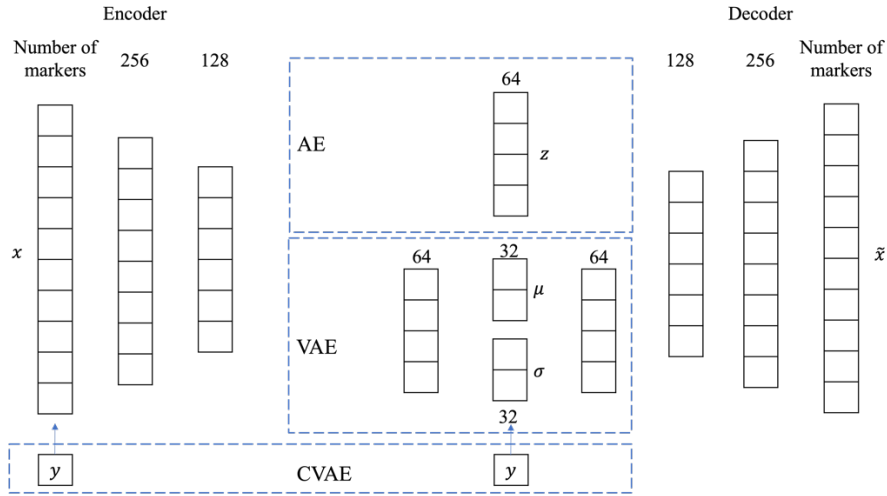

**Fig. S11 | Structure of the VAEs used in SyNG-BTS.** For autoencoder (AE), the encoder consists of two hidden layers with 256 and 128 nodes, and the low-dimensional representation has a dimension of 64; the decoder has a symmetric structure of hidden layers. In the case of the VAE, both the encoder and decoder have an additional hidden layer with 64 nodes, and the low-dimensional representation '$z$' is a 32-dimensional Gaussian vector with mean '$\mu$' and covariance matrix '$\sigma^2 I$'. The Conditional VAE (CVAE) shares the same encoder and decoder structure as the VAE, with the only difference being that the input of both the encoder and decoder includes one additional node for the sample group labels.

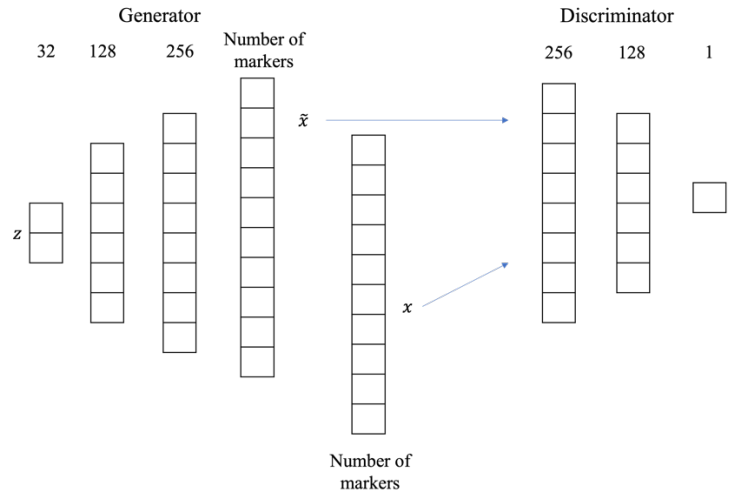

**Fig. S12 | Structure of the GANs used in SyNG-BTS.** GAN, WGAN and WGANGP share the same structure for both the generator and discriminator. The generator starts with a 32-dimensional random Gaussian noise and consists of two hidden layers with 128 and 256 nodes, respectively. The discriminator takes generated samples $\tilde{x}$ and real samples $x$ as input, featuring two hidden layers with 256 and 128 nodes and an output layer with only 1 node. The distinction among the three GAN model variants lies in their loss functions.

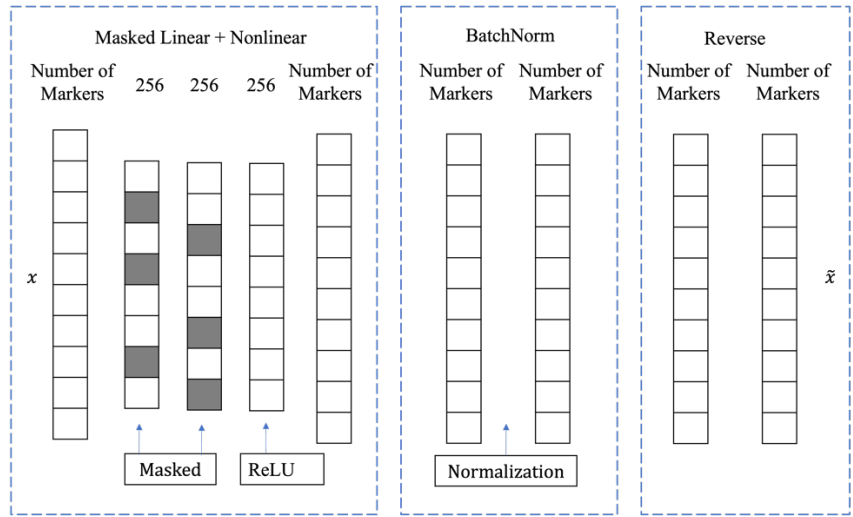

**Fig. S13 | Structure of the Flow-based models used in SyNG-BTS.** MAF, GLOW and RealNVP each consists of three Neural Network components: (i) *Masked Linear + Nonlinear*: This component includes hidden layers consisting of 256 nodes, employs a masking strategy with 30% of the connections masked, and utilizes the ReLU activation function for non-linear transformations; (ii) *Batch Normalization*: Implemented following each layer, batch normalization standardizes the output of the previous layer by re-centering and re-scaling, improving training stability and speed; and (iii) *Reverse*: This component is tasked with computing the inverse function, featuring hidden layers that also contain 256 nodes. This aspect of the architecture ensures the model's invertibility, a critical aspect of Flow-based models that enables efficient computation of likelihoods and straightforward sampling.